\RequirePackage{fix-cm}
\documentclass[twocolumn,epjc3]{svjour3}
\smartqed  
\RequirePackage{graphicx,amsmath}
\usepackage{amssymb,latexsym,amsgen}
\usepackage{cite}
\usepackage[colorlinks=true,
            linkcolor=blue,
            citecolor=blue,
            urlcolor=blue]{hyperref}

\usepackage{footmisc}

\journalname{Eur. Phys. J. C}

\begin{document}

\title{Resonant leptogenesis in inverse see-saw framework with modular $S_4$ symmetry}

\author{Abhishek\thanksref{e1,addr1}
        \and
        V.~Suryanarayana~Mummidi\thanksref{e2,addr1}
}

\thankstext{e1}{e-mail: 413120051@nitt.edu}
\thankstext{e2}{e-mail: venkata@nitt.edu}

\institute{Department of Physics, National Institute of Technology,\\
Tiruchirappalli-620015, India \label{addr1}
}

\date{Received: date / Accepted: date}

\maketitle

\begin{abstract}
We introduce a lepton mass generation and flavor mixing model, realized through a (2,3) inverse seesaw structure based on modular \( S_4 \) symmetry. The model employs modular forms to construct the lepton Yukawa couplings, significantly simplifying the framework by reducing redundant parameters. A detailed numerical analysis demonstrates consistency with current neutrino oscillation data, yielding specific outputs for the mixing angles and CP-violating phases. The Dirac CP phase is predicted to lie near $\delta_{\rm CP} \approx \pm 90^\circ$, corresponding to near-maximal leptonic CP violation. The total neutrino mass lies within $\sum m_\nu \approx 0.0587\text{--}0.0924$ eV, and the effective Majorana mass $|m_{ee}| \approx (0.002\text{--}0.02)$ eV, within reach of upcoming neutrinoless double beta decay experiments such as nEXO and AMoRE-II. The model also remains consistent with current bounds on charged lepton flavor violating processes from MEG and BaBar. We further explore resonant leptogenesis enabled by quasi-degenerate heavy neutrino states and show that the observed baryon asymmetry of the universe can be successfully generated in this scenario. The combined treatment of low-energy observables and high-scale baryogenesis demonstrates the predictivity and testability of the modular \( S_4 \)-based ISS(2,3) framework.
\end{abstract}
\tableofcontents
\section{Introduction}

Neutrino oscillation measurements confirm that neutrinos are not massless and exhibit flavor mixing \cite{SuperK1998, SNO2002}. These findings prove physics beyond the Standard Model (SM). The precise dynamics responsible for endowing neutrinos with mass and shaping the lepton mixing matrix have yet to be fully established. To overcome these problems, we propose to extend the SM  with a mechanism that produces tiny neutrino masses while adding discrete flavor symmetries to restrict the model's flavor structure~\cite{King_2013}. Many extensions have been proposed thus far, but the inverse seesaw mechanism is an attractive framework, it accommodates sub-eV neutrino masses without requiring ultra-high mass scales~\cite{Nomura:2021, Mohapatra:1986,Mohapatra:1986b,Bernabeu:1987,Gavela:2009,Parida:2010,Garayoa:2007,Abada:2014,Law:2013,Nguyen:2020,Deppisch:2005,Arina:2008,Dev:2010,Malinsky:2009,Hirsch:2010,Blanchet:2010,Dias:2012,Agashe:2019,Gautam:2020,Zhang:2021}. The inverse seesaw differs from the conventional type-I scenario~\cite{Brdar:2019,Schechter:1982,MINKOWSKI1977421} by enabling TeV-scale heavy neutrinos via the introduction of singlet fermions and a softly broken lepton number, thus avoiding unnaturally small Yukawa couplings. This makes the inverse seesaw phenomenologically rich and potentially testable at current and future colliders. 

To introduce additional constraints on the neutrino mass matrix, we formulate the inverse seesaw mechanism within a modular flavor symmetry paradigm, where the underlying symmetry group is \( \Gamma_4 \), known to be isomorphic to the permutation group \( S_4 \)~\cite{Feruglio2017,Kobayashi:2019,Penedo:2019,Wang:2020,Zhang:2021}. Modular symmetries provide a framework for flavor model building, where Yukawa couplings arise as modular forms determined by the symmetry, and the complicated flavon vacuum alignment of traditional models is replaced by the complex modulus $\tau$, whose vacuum expectation value governs both mass hierarchies and mixing patterns. In addition to accurately capturing the observed lepton-mixing pattern and neutrino mass splittings, the framework significantly limits the number of independent parameters, enhancing its predictive power. Moreover, the modular group structure connects particle physics with the mathematics of modular forms, offering a minimal and highly constrained framework, and has become a focus of recent model-building efforts.

In this work, we construct a minimal inverse seesaw model based on $S_4$ modular symmetry and examine its effects on both low and high energy observables. To test the viability of our model, we analyze up-to-date global neutrino oscillation data, including the CP phase \( \delta_{\rm CP}\), mixing angles, and mass-squared splittings. Modular weight assignments and group representations are selected to replicate the observed charge lepton mass hierarchy while preserving the predictivity of the model in the neutrino sector. Our analysis shows that the model aligns well with the current oscillation data, remaining within the confidence intervals \( 3\sigma \). In addition to neutrino oscillation observables, the framework also allows the study of charged lepton flavor violation (LFV) and neutrinoless double beta decay ($0\nu\beta\beta$). Processes such as $\mu \to e\gamma$, $\tau \to \mu\gamma$ and $\tau \to e\gamma$ are naturally expected in the inverse seesaw due to the presence of additional singlet states, while contributions to $0\nu\beta\beta$ can arise from both light and heavy neutrino exchange. These phenomena provide complementary low-energy tests of the model, and we study their implications within this framework. This setup ensures that predictions for LFV rates, the effective Majorana mass $m_{ee}$ and leptogenesis are not independent but originate from the same Yukawa structures determined by the modular symmetry.

This study implements the ISS(2,3) mechanism with-in a modular $S_4$ flavor symmetry and examines its phenomenological consequences. While modular symmetries have been studied with other seesaw mechanisms, the inverse seesaw has been explored without such symmetry constraints. Here, we bring these two approaches together and analyze the resulting implications consistently across neutrino oscillations, lepton flavor violation, neutrinoless double beta decay and leptogenesis.

A particularly appealing feature of the inverse seesaw model is its compatibility with low-scale leptogenesis, especially within the framework of resonant leptogenesis~\cite{Pilaftsis1997,Iso:2014,Dev:2014,Aoki:2015,Dev:2018,Asaka:2019,Chakraborty:2021azg}. In our setup, two of the heavy neutrinos form a quasi-degenerate pair with small mass difference - naturally arising from the structure of ISS(2,3) model - can lead to a resonant enhancement of CP asymmetry in their decays. By studying the deacays of quasi-degenarate heavy right handed neutrinos, we investigate whether the baryon-to-photon ratio inferred from cosmological observations can be successfully produced via thermal leptogenesis. The CP-violating asymmetry originates from the quantum interference between leading-order and radiative-level contributions to decay processes of the heavy neutrinos. It is numerically evaluated using Yukawa structures determined by the underlying modular symmetry. We solve the Boltzmann equations that govern the thermal evolution of the heavy neutrino densities and the lepton asymmetry, considering the relevant decay and washout processes. Our analysis shows that the model can generate the correct baryon asymmetry within a viable range of parameters, without requiring extremely high-scale physics.

The paper is organized as follows. The theoretical structure of the model, which includes the inverse seesaw mechanism integrated into modular $S_4$ symmetry, is presented in Section~\ref{sec:Model}. Section~\ref{sec:NA} contains the numerical fitting of neutrino oscillation parameters, followed by a phenomenological analysis divided into two parts: one addressing LFV processes and the other exploring neutrinoless double beta decay signatures. In Section~\ref{sec:RL}, we describe the model's resonant leptogenesis mechanism and derive the relevant CP asymmetries. Section~\ref{sec:NAL} covers the numerical study of leptogenesis via Boltzmann equations, analyzing the generation of baryon asymmetry. Section~\ref{sec:C} concludes with a summary of our results.

\section{Model}
\label{sec:Model}
The model is built upon the modular symmetry \( S_4 \) and implements the inverse seesaw ISS(2,3) configuration. The proposed framework, which is regulated by the flavor symmetry $S_4 \times k_I$, where $k_I$ stands for modular weights, adds three sterile fermions $S_i$ and pair of right-handed neutrino fields $N_R$ to the SM. In Table~\ref{tab:fields}, the particle content and transformation properties are compiled, with the lepton doublets $L$ transforming as an $S_4$ triplet, right-handed charged leptons $l_R$ as singlets ($1$ or $1'$), and the right-handed neutrinos $N_R$ as an $S_4$ doublet. Higgs doublets $H_u$, $H_d$, and a gauge singlet $\varphi$ make up the scalar sector; they are all invariant under $S_4$. The modular weights are assigned such that all fermions carry $k_I = -2$, while the Yukawa couplings have $k_I = 2$, as shown in Table~\ref{tab:yukawa}.

\begin{table}[h]
\centering
\caption{The model's particle composition and charges}
\label{tab:fields}
\begin{tabular}{cccccccc}
\hline
Field & $L$ & $l_{R_1}^c,l_{R_2}^c,l_{R_3}^c$ & $N_R^c$ & $S_i$ & $H_u$ & $H_d$ & $\varphi$ \\
\hline
$SU(2)_L$ & 2 & 1 & 1 & 1 & 2 & 2 & 1 \\
$S_4$ & 3 & $1'$, $1$, $1'$ & 2 & 3 & 1 & 1 & 1 \\
$k_I$ & $-2$ & 0, $-2$, -2 & $-2$ & $-2$ & 0 & 0 & 0 \\
\hline
\end{tabular}
\end{table}

\begin{table}[h]
\centering
\caption{Yukawa interactions dictate the assignment of modular weights.}
\label{tab:yukawa}
\begin{tabular}{cccc}
\hline
Yukawa coupling &  $Y_{3'}$ & $Y_3^{(4)}$ & $Y_{3'}^{(4)}$ \\
\hline
$S_4$ & $3'$ & 3 & $3'$ \\
\hline
$k_I$ & 2 & 4 & 4 \\
\hline
\end{tabular}
\end{table}

The Yukawa couplings $Y$ are modular forms of weight 2 transforming as an $S_4$ triplet, with higher-weight forms $Y^{(4)}_3$ constructed from their products. This structure significantly constrains the possible interactions in the superpotential while maintaining predictive power. The lepton sector is described by
\begin{equation}
    W_{\text{lepton}} = W_l + W_D + W_{NS} + W_S,
\end{equation}
where $W_l$ generates charged lepton masses, $W_D$ mediates Dirac neutrino masses, $W_{NS}$ governs the mixing between $N_R$ and $S_i$, and $W_S$ provides Majorana masses for the sterile fermions.

The following terms give rise to the charged lepton mass matrix:
\begin{equation}
\begin{split}
W_l = \alpha ({L}l_{R_1}^c)_{3'} (H_d)_1 Y_{3'} 
+ \beta ({L}l_{R_2}^c)_3 (H_d)_1 Y^{(4)}_3 \\
+ \gamma ({L}l_{R_3}^c)_{3'} (H_d)_1 Y^{(4)}_{3'}.
\end{split}
\end{equation}

yielding the explicit form
\begin{equation}
M_L = v_d \begin{pmatrix}
\alpha Y_3 & -2\beta Y_2 Y_3 & 2\gamma Y_1 Y_3 \\
\alpha Y_5 & \beta (\sqrt{3} Y_1 Y_4 + Y_2 Y_5) & \gamma (\sqrt{3} Y_2 Y_4 - Y_1 Y_5) \\
\alpha Y_4 & \beta (\sqrt{3} Y_1 Y_5 + Y_2 Y_4) & \gamma (\sqrt{3} Y_2 Y_5 - Y_1 Y_4)
\end{pmatrix},
\end{equation}
We define \( v_d \) as the vev of \( H_d \), responsible for giving mass to charged leptons. In parallel, the Dirac neutrino mass matrix arises from an \( S_4 \)-invariant interaction among the lepton doublets \( L \), right-handed neutrinos \( N_R \) and the up-type Higgs \( H_u \), structured by the modular forms \( Y_{3,3'}^{(4)} \) of weight 4.
  The superpotential takes the form:
\begin{equation}
W_D = \alpha_D ({L} N_R^c)_3 H_u Y_3^{(4)} + \beta_D ({L} N_R^c)_{3'} H_u Y_{3'}^{(4)},
\label{eq:2.5}
\end{equation}
where $\alpha_D$ and $\beta_D$ are coupling constants. 

The resulting $3 \times 2$ Dirac mass matrix exhibits a constrained structure due to $S_4$ tensor products:
\begin{equation}
M_D = v_u
\resizebox{0.85\columnwidth}{!}{$
\begin{pmatrix}
-2\alpha_D a & -2\beta_D b \\
-\alpha_D\!\left(\tfrac{\sqrt{3}}{2} e + \tfrac{1}{2} d\right)
 + \beta_D\!\left(\tfrac{3}{2} d - \tfrac{\sqrt{3}}{2} e\right) &
 \alpha_D\!\left(\tfrac{3}{2} f + \tfrac{\sqrt{3}}{2} c\right)
 + \beta_D\!\left(\tfrac{\sqrt{3}}{2} c - \tfrac{1}{2} f\right) \\
-\alpha_D\!\left(\tfrac{\sqrt{3}}{2} f + \tfrac{1}{2} c\right)
 + \beta_D\!\left(\tfrac{3}{2} c - \tfrac{\sqrt{3}}{2} f\right) &
 \alpha_D\!\left(\tfrac{3}{2} e + \tfrac{\sqrt{3}}{2} d\right)
 + \beta_D\!\left(\tfrac{\sqrt{3}}{2} d - \tfrac{1}{2} e\right)
\end{pmatrix}$}
\end{equation}

where $v_u$ is the vev of $H_u$ and the parameters $a$ through $f$ represent specific combinations of modular forms:
\begin{align*}
      a &= Y_2 Y_3, \quad b = Y_1 Y_3, \quad c = Y_2 Y_4, \\
      d &= Y_2 Y_5, \quad e = Y_1 Y_4, \quad f = Y_1 Y_5.
\end{align*}
This matrix structure reflects the branching rules of $S_4$ representations when contracting the triplet ${L}$ with the doublet $N_R$, where each entry is explicitly calculable in terms of the underlying modular forms. All transformations of the Yukawa couplings under the modular \( S_4 \) symmetry, along with their representation structure and modular forms, are summarized in Appendix~B. The interaction between the right-handed neutrinos \( N_R \) and the sterile fermions \( S_i \) is given by: 
\begin{equation}
W_{NS} = \alpha_{NS} (Y_3^{(4)} N_R)_3 (S)_3 \varphi + \beta_{NS} (Y_{3'}^{(4)} N_R)_3 (S)_{3} \varphi
\end{equation}
where $\alpha_{NS}$ and $\beta_{NS}$ are free parameters and $v_{NS}$ is the vev of the singlet scalar $\varphi$. This generates the $2 \times 3$ mass matrix:
\begin{equation}
M_{NS} = v_{NS}
\begin{pmatrix}
a_{11} & a_{12} & a_{13} \\
a_{21} & a_{22} & a_{23}
\end{pmatrix},
\label{eq:MNS_general}
\end{equation}
where
\begin{align}
a_{11} &= -2\alpha_{NS} a, \nonumber \\
a_{12} &= -\tfrac{\alpha_{NS}}{2}(\sqrt{3} e + d)
         + \tfrac{\sqrt{3}}{2}\beta_{NS}(\sqrt{3} d - e), \nonumber \\
a_{13} &= -\tfrac{\alpha_{NS}}{2}(\sqrt{3} f + c)
         + \tfrac{\sqrt{3}}{2}\beta_{NS}(\sqrt{3} c - f), \nonumber \\
a_{21} &= -2\beta_{NS} b, \nonumber \\
a_{22} &= \tfrac{\sqrt{3}}{2}\alpha_{NS}(\sqrt{3} f + c)
         + \tfrac{\beta_{NS}}{2}(\sqrt{3} c - f), \nonumber \\
a_{23} &= \tfrac{\sqrt{3}}{2}\alpha_{NS}(\sqrt{3} e + d)
         + \tfrac{\beta_{NS}}{2}(\sqrt{3} d - e).
\label{eq:MNS_elements}
\end{align}

whose structure is fixed by the tensor products of $S_4$ representations and the modular form combinations $Y_i Y_j$. 

The Majorana mass for the sterile states $S_i$ is generated through their coupling to the weight-4 modular forms $Y_3^{(4)}$, with the superpotential given by:
\begin{equation}
W_S = \mu_0 \left({S} S\right)_3 Y_3^{(4)}
\end{equation}
where \( \mu_0 \) acts as the parameter that controls the extent of violation of the lepton number. Note that the Majorana term $({S} S)_{3'} Y_{3'}^{(4)}$ is forbidden, since the $3'$-dimensional representation is antisymmetric. 
The resulting $3 \times 3$ Majorana mass matrix exhibits a distinctive symmetric structure:
\begin{equation}
M_S = \mu_0 \begin{pmatrix}
0 & -(\sqrt{3}f + c) & \sqrt{3}e + d \\
-(\sqrt{3}f + c) & 2a & 0 \\
\sqrt{3}e + d & 0 & -2a
\end{pmatrix}
\end{equation}
This mass matrix significantly contributes to inverse seesaw mechanism, where the light neutrino masses are suppressed by the small ratio $M_S/M_{NS}^2$.  In the $(\nu_L, N_R, S_i)$ basis, the entire neutrino mass matrix arranges itself in an inverse seesaw pattern:
\begin{equation}
\mathcal{M}_\nu = \begin{pmatrix}
0 & M_D & 0 \\
M_D^T & 0 & M_{NS} \\
0 & M_{NS}^T & M_S
\end{pmatrix},
\label{eq:full_matrix}
\end{equation}
At energy scales below the $N_R$ and $S_i$ masses, the mass matrix of light neutrinos emerges through the inverse seesaw mechanism~\cite{Zhang:2021}:
{\footnotesize
\begin{multline}
M_\nu = -M_D \bigl(M_{NS}^T M_{NS}\bigr)^{-1} 
M_{NS}^T M_S M_{NS} 
\\
\times \bigl(M_{NS}^T M_{NS}\bigr)^{-1} M_D^T,
\label{eq:seesaw}
\end{multline}
}

\begin{figure*}[htbp]
    \centering
    \begin{minipage}{0.45\textwidth}
        \centering
        \includegraphics[width=\linewidth]{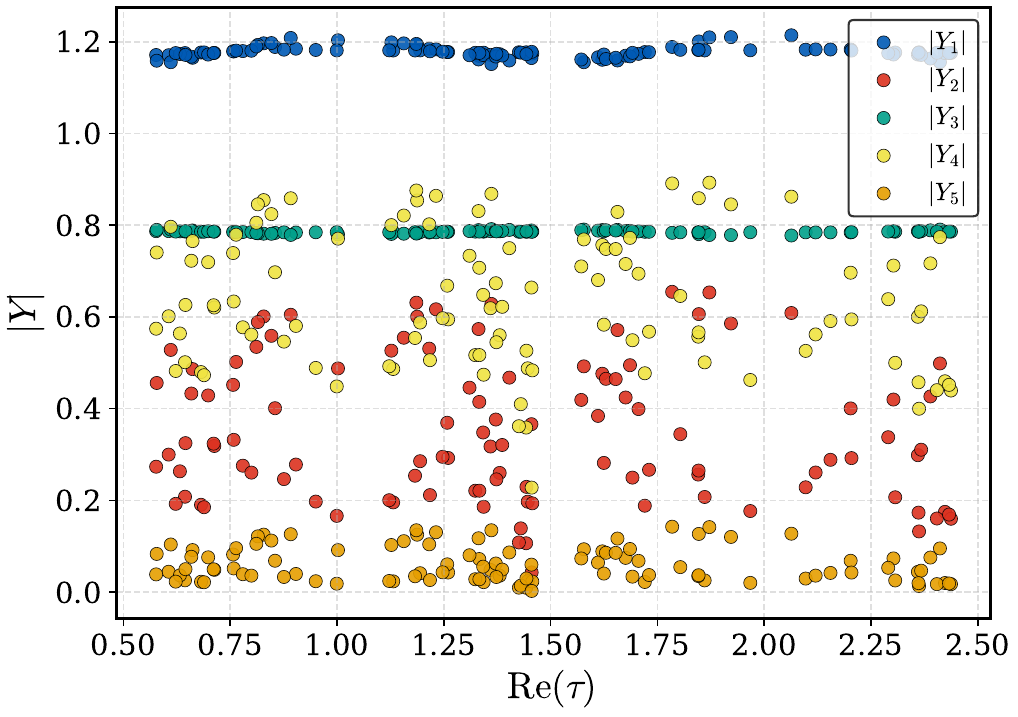}
    \end{minipage}
    \hfill
    \begin{minipage}{0.45\textwidth}
        \centering
        \includegraphics[width=\linewidth]{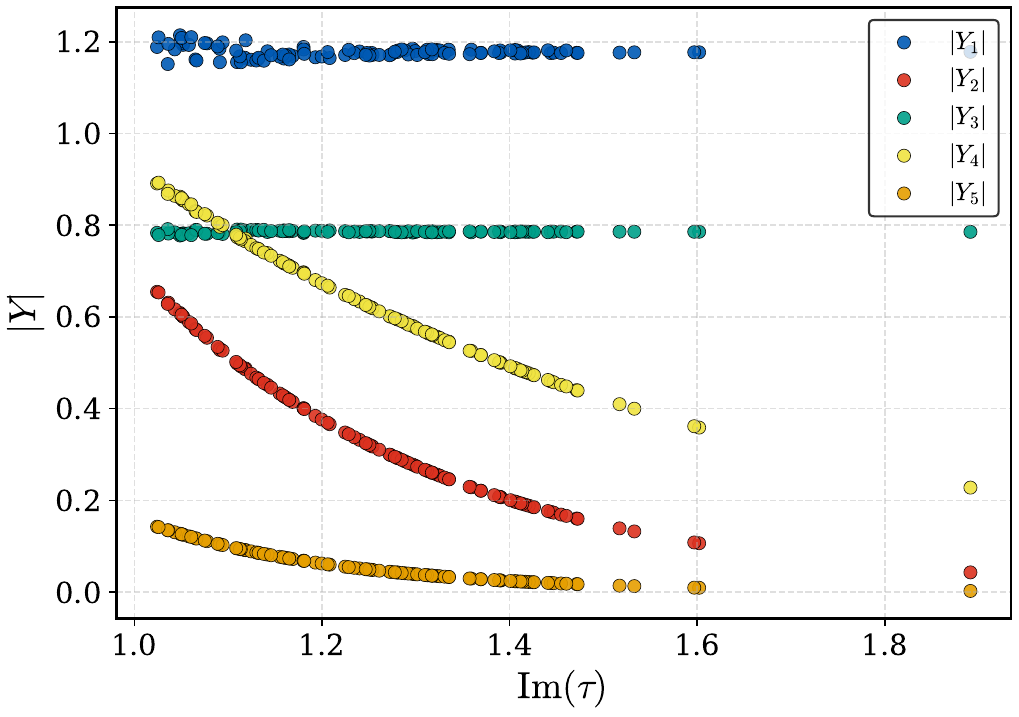}
    \end{minipage}
    \caption{
    Correlation between Yukawa coupling magnitude and the real (left) and imaginary (right) components of the complex modulus \( \tau \). The plots highlight how the modular forms, controlled by \( \tau \), shape the Yukawa structure in the \( S_4 \)-based inverse seesaw model.
    }
    \label{fig:combined_figures}
\end{figure*}

Before proceeding to the numerical analysis, we brief-ly comment on the supersymmetry-breaking scale and its mediation to the visible sector. Although the analysis in this work is performed in the supersymmetric limit, modular construction $S_4$ can be consistently embedded in a framework where supersymmetry breaking is communicated to the visible sector through higher-dimensional operators. Breaking may be parameterized by a spurion superfield $X = \theta^2 F$, whose vacuum expectation value generates soft terms of order $m_{\text{SUSY}} \simeq {F}/{M_{\text{mess}}}$,
with $M_{\text{mess}}$ denoting the characteristic messenger scale of SUSY-breaking mediation.
For the parameter range relevant to our model, the heavy neutrino masses lie around $M_N \sim 10^{4\text{--}5}~\mathrm{GeV}$. A consistent and phenomenologically safe hierarchy,
\[
m_{\text{SUSY}} \sim 10^{6\text{--}8}~\mathrm{GeV}, \quad M_{\text{mess}} \sim 10^{14\text{--}16}~\mathrm{GeV},
\]
yields a ratio $m_{\text{SUSY}}/M_{\text{mess}} \leq 10^{-8}$.
Such suppression ensures that supersymmetry-induced corrections to the modular Yukawa couplings remain negligible, preserving the predictive modular texture of the lepton sector~\cite{Criado:2018ths}.
In this regime, all superpartners decouple well before the leptogenesis epoch, and the effective low-energy description employed throughout this paper remains valid.

\section{Numerical analysis}
\label{sec:NA}
\subsection{Neutrino Oscillation Data Fit}

We numerically evaluate the model’s predictions to ensure consistency with current neutrino oscillation data. This is achieved through the PMNS matrix, which encapsulates the non-trivial mixing between the neutrino and charged lepton sectors. Since both the the neutrino mass matrix and charged lepton mass matrix are not initially in diagonal form, we must diagonalize them to extract the physical mixing parameters. The process of bringing the charged lepton mass matrix to a diagonal form $M_\ell$ is as follows:
\begin{equation}
U_\ell^\dagger M_\ell M_\ell^\dagger U_\ell = \mathrm{diag}(m_e^2, m_\mu^2, m_\tau^2),
\end{equation}
where $U_\ell$ is the unitary matrix that transforms the charged left-handed lepton fields into their mass eigenstates.

Similarly, the neutrino mass matrix $M_\nu$ is diagonalized as:
\begin{equation}
U_\nu^\dagger M_\nu M_\nu^\dagger U_\nu = \mathrm{diag}(m_1^2, m_2^2, m_3^2),
\end{equation}
where the diagonalizing matrix $U_\nu$ transforms the neutrino mass matrix into its mass basis, and the eigenvalues $m_i^2$ give the physical neutrino masses. Next, we obtain the lepton mixing matrix, which is also referred to as the PMNS matrix:
\begin{equation}
U_{\mathrm{PMNS}} = U_\ell^\dagger U_\nu.
\end{equation}

The structure of the PMNS matrix involves three mixing angles and a single Dirac CP-violating phase $(\delta_{\mathrm{CP}})$ and two additional Majorana phases $(\alpha_{21}, \alpha_{31})$. The standard parametrization is given by:
\begin{equation}
U_{\mathrm{PMNS}} = \resizebox{0.38\textwidth}{!}{$
\begin{pmatrix}
c_{12} c_{13} & s_{12} c_{13} & s_{13} e^{-i\delta_{\mathrm{CP}}} \\
-s_{12} c_{23} - c_{12} s_{23} s_{13} e^{i\delta_{\mathrm{CP}}} & 
c_{12} c_{23} - s_{12} s_{23} s_{13} e^{i\delta_{\mathrm{CP}}} & 
s_{23} c_{13} \\
s_{12} s_{23} - c_{12} c_{23} s_{13} e^{i\delta_{\mathrm{CP}}} & 
-c_{12} s_{23} - s_{12} c_{23} s_{13} e^{i\delta_{\mathrm{CP}}} & 
c_{23} c_{13}
\end{pmatrix}
\cdot
P_\nu$},
\end{equation}
where \( P_\nu = \mathrm{diag}(1, e^{i\alpha_{21}/2}, e^{i\alpha_{31}/2}) \), and for $i,j = 1,2,3$, $c_{ij} = \cos \theta_{ij}$ and $s_{ij} = \sin \theta_{ij}$.

Using the components of the PMNS matrix, the lepton sector’s mixing angles can be explicitly determined via:
\begin{align}
\sin^2\theta_{13} &= |U_{e3}|^2, \\
\sin^2\theta_{12} &= \frac{|U_{e2}|^2}{1 - |U_{e3}|^2}, \\
\sin^2\theta_{23} &= \frac{|U_{\mu 3}|^2}{1 - |U_{e3}|^2}.
\end{align}

\begin{table}[t]
\centering
\caption{The model parameters have been adjusted to align with the observed neutrino oscillation data~\cite{Esteban:2024}.}
\renewcommand{\arraystretch}{1.2}
\setlength{\tabcolsep}{12pt}

\begin{tabular}{c c c} 
\hline
\textbf{Parameter} & \textbf{Best-fit $\pm$ 1$\sigma$} & \textbf{3$\sigma$ range}  \\ 
\hline

$\sin^2 \theta_{12}$ & $0.307^{+0.012}_{-0.011}$ & $0.275 - 0.345$  \\  
$\sin^2 \theta_{23}$ & $0.561^{+0.012}_{-0.015}$ & $0.430 - 0.596$  \\  
$\sin^2 \theta_{13}$ & $0.02195^{+0.00054}_{-0.00058}$ & $0.02023 - 0.02376$  \\
$\Delta m^2_{21}\ [10^{-5}\ \mathrm{eV}^2]$ & $7.49^{+0.19}_{-0.19}$ & $6.92 - 8.05$  \\  
$\Delta m^2_{31}\ [10^{-3}\ \mathrm{eV}^2]$ & $2.534^{+0.025}_{-0.023}$ & $2.463 - 2.606$  \\
  
\hline
\end{tabular}
\label{tab:3}
\end{table}

\begin{figure*}[htbp]
    \centering
    \begin{minipage}{0.49\linewidth}
        \centering
        \includegraphics[width=\linewidth]{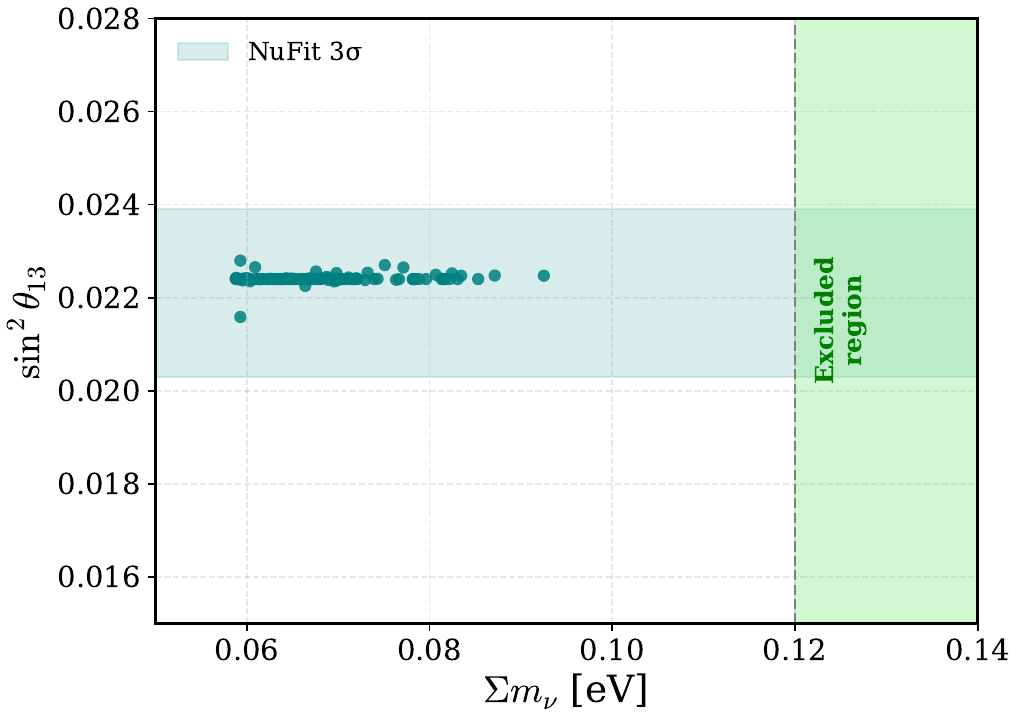}
    \end{minipage}
    \hfill
    \begin{minipage}{0.49\linewidth}
        \centering
        \includegraphics[width=\linewidth]{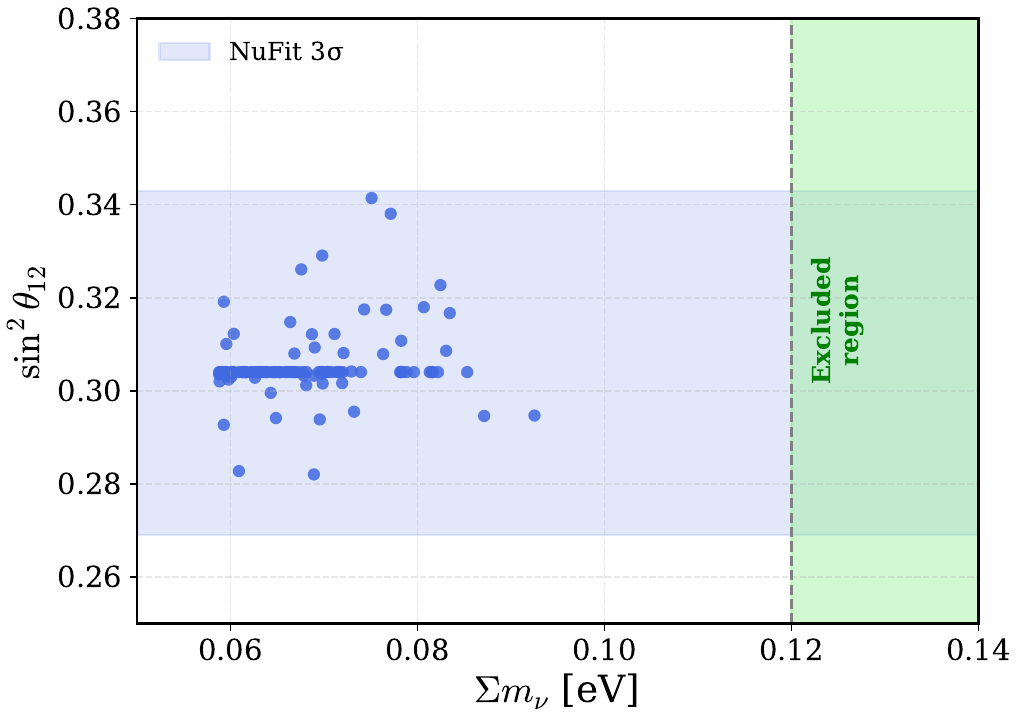}
    \end{minipage}
    
    \vspace{0.5cm}
    \begin{minipage}{0.49\linewidth}
        \centering
        \includegraphics[width=\linewidth]{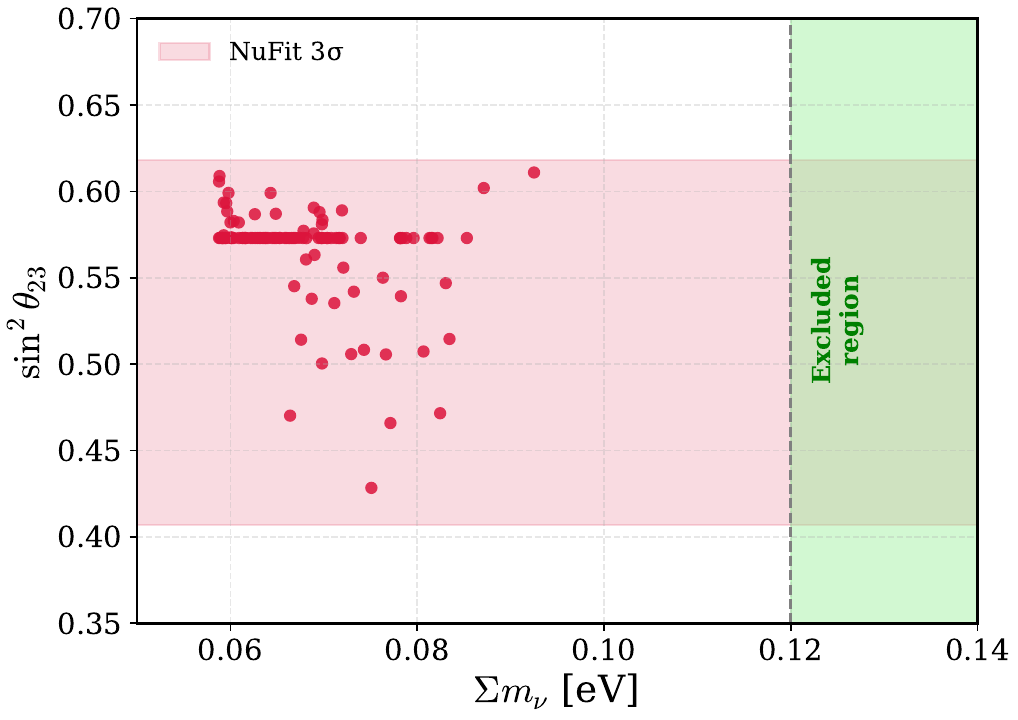}
    \end{minipage}
    
    \caption{
    The relationship between the mixing angles $\theta_{13}$, $\theta_{12}$, $\theta_{23}$ and the sum of neutrino masses \( \sum m_\nu \) (eV). 
    }
    \label{fig:theta13_sum}
\end{figure*}

To quantify CP violation in the lepton sector, we use the Jarlskog invariant:
\begin{equation}
J_{\mathrm{CP}} = \text{Im}[U_{e1} U_{\mu 2} U_{e2}^* U_{\mu 1}^*] = s_{12} c_{12} s_{23} c_{23} s_{13} c_{13}^2 \sin\delta_{\mathrm{CP}},
\end{equation}
and the Majorana phases are probed through the following rephasing-invariant combinations:
\begin{align}
I_1 &= \text{Im}[U_{e1}^* U_{e2}]^2 = c_{12} s_{12} c_{13} \sin\alpha_{21}, \\
I_2 &= \text{Im}[U_{e1}^* U_{e3}]^2 = c_{12} s_{13} c_{13} \sin(\alpha_{31} - 2\delta_{\mathrm{CP}}).
\end{align}
To identify parameter points that match the oscillation data for normal ordering in Table~\ref{tab:3}, we perform an random scan over the model's free parameters within the ranges specified in Table~\ref{tab:params}.

\begin{table}[b]
\centering
\caption{Model parameter ranges that are used in the numerical analysis.}
\renewcommand{\arraystretch}{1.2}
\begin{tabular}{|c|c|}
\hline
\textbf{Parameter} & \textbf{Range} \\
\hline
$\text{Re}[\tau]$ & $[0.5,\ 2.5]$ \\
$\text{Im}[\tau]$ & $[1,\ 3]$ \\
$\alpha, \beta, \gamma$   & $[0.5,\ 2]$ \\
$\alpha_D, \beta_D$        & $[0.02 ,\ 0.06 $ \\
$v_{NS}$ [GeV]    & $[ 3\times10^4,\  8\times10^4 ]$ \\
$\alpha_{NS}, \beta_{NS}$     & $[0.3,\ 0.6]$ \\
 
$\mu_0$ [GeV]     & $[10^{-3},\ 10^3]$ \\
\hline
\end{tabular}

\label{tab:params}
\end{table}
These parameter ranges are chosen to ensure that the model remains consistent with low-energy phenome-nology and to allow for viable texture structures in the mass matrices. The Higgs doublets \( H_u \) and \( H_d \) acquire vevs related to the electroweak scale by \( v_H^2 = v_u^2 + v_d^2 \), where \( v_H = 246 \) GeV. In our numerical analysis, we work in the approximation where the ratio of these vevs is fixed to \( \tan\beta = 5 \)~\cite{Antusch:2013jca, Kashav:2021zir}.

The results of our numerical scan are presented in the following correlation plots, which illustrate the relationships between various neutrino observables obtained from our parameter scan. Figures~\ref{fig:combined_figures} illustrate how the modular Yukawa couplings \( Y_i(\tau) \) vary with changes in both components of the complex modulus \( \tau \). Since the $Y_i$ are modular forms of weight 4, they exhibit non-trivial but controlled variations across the domain. These patterns play a key role in determining the textures and hierarchies of the lepton mass matrices.

Figures~\ref{fig:theta13_sum} illustrate how the predicted lepton mixing angles vary with respect to the total neutrino mass \( \sum m_\nu \) within the context of the proposed framework. The resulting ranges for the neutrino mixing parameters, $\sin^2\theta_{12} \approx 0.30\text{--}0.32$, $\sin^2\theta_{13} \approx 0.021\text{--}0.0225$, and $\sin^2\theta_{23} \approx 0.55\text{--}0.60$, fall within the \( 3\sigma \) intervals reported in the NuFIT 6.0 global analysis. The total neutrino mass is predicted to lie within:
\[
\sum m_\nu \in [0.0587,\ 0.0924]~\text{eV},
\]
well below the Planck upper bound of $0.12$~eV~\cite{Planck2018Parameters}. These results confirm that the model accommodates both oscillation data and cosmological constraints.
\begin{figure}[t]
    \centering
    \includegraphics[width=0.95\linewidth]{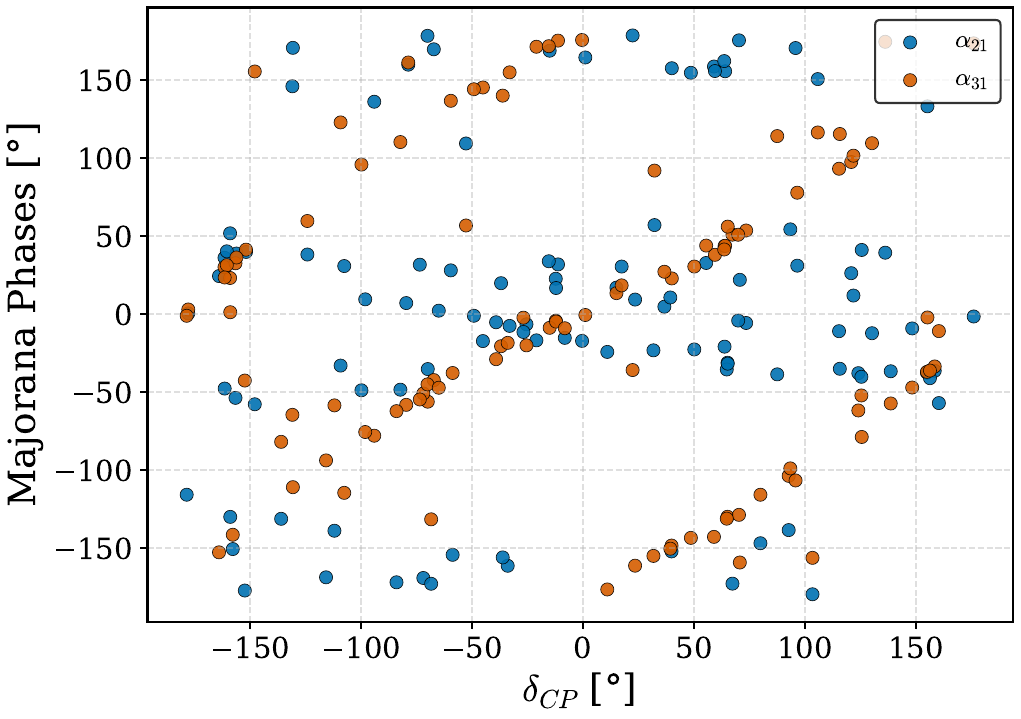}
   \caption{
Interplay between the Dirac CP-violating phase \( \delta_{\rm CP} \) and the two Majorana phases \( \alpha_{21} \) and \( \alpha_{31} \).
}
    \label{fig:alpha21_dcp}
\end{figure}
Figures~\ref{fig:alpha21_dcp}  illustrates the relation between the Majorana CP phases ($\alpha_{21}$, $\alpha_{31}$) and the Dirac CP phase $\delta_{\mathrm{CP}}$. We find $\delta_{\rm CP} \approx \pm 90^\circ$, with Majorana phases $\alpha_{21} \in [-100^\circ, +100^\circ]$ and $\alpha_{31} \in [-120^\circ, +120^\circ]$ exhibiting broad but symmetric distributions. This suggests a strong prediction of near-maximal CP violation, that could be probed in upcoming experiments such as DUNE and Hyper-Kamiokande~\cite{Song_2021}.

\begin{figure*}[t]
    \centering
    \includegraphics[width=0.48\textwidth]{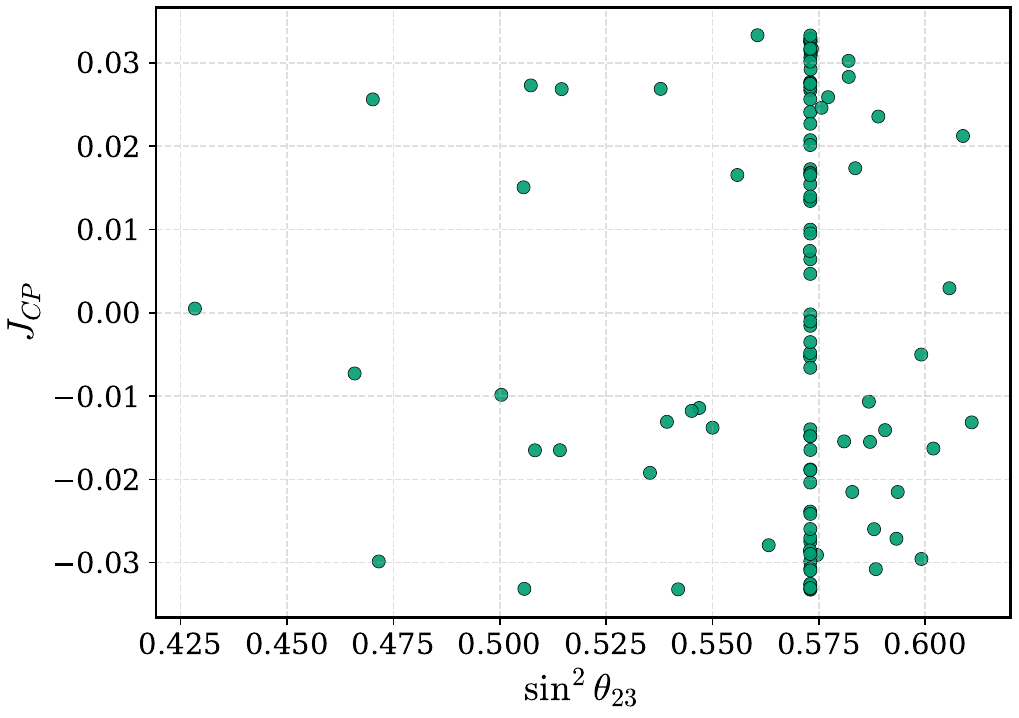}
    \hfill
    \includegraphics[width=0.48\textwidth]{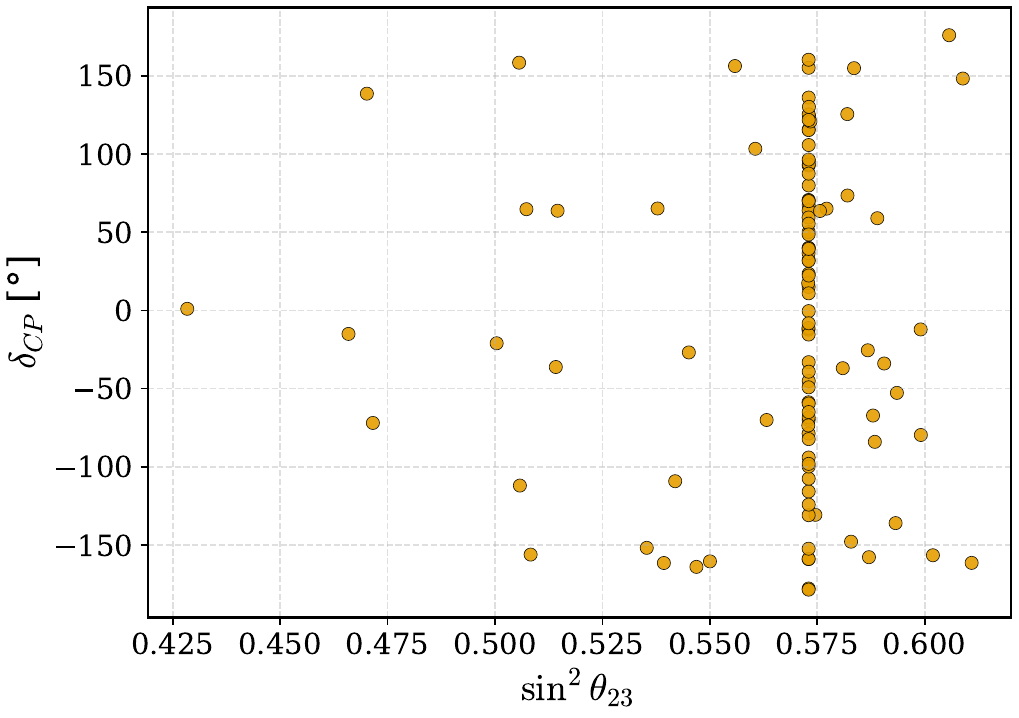}
    \caption{
        Dependence of the Jarlskog invariant ($J_{CP}$) on $\sin^2\theta_{23}$ and the Dirac phase \( \delta_{\rm CP} \).
    }
    \label{fig:5}
\end{figure*}

We show the expected values of the Jarlskog invariant in Figure~\ref{fig:5} (left panel), throughout the range:
\[
J_{\mathrm{CP}} \in [-0.03, 0.03],
\]
with respect to $\sin^2\theta_{23}$. The values are in the expected range for lepton sector CP violation and highlight the constrained nature of CP-violating parameters in this model. In Figure~\ref{fig:5} (Right panel), demonstrates the strong correlation between $\delta_{\mathrm{CP}}$ and $\sin^2\theta_{23}$, with both parameters simultaneously constrained to their experimental ranges.

To assess the robustness of the obtained best-fit solution, we have performed a detailed $\chi^2$-sensitivity analysis by varying each model parameter individually while keeping all others fixed at their best-fit values. The resulting one-dimensional $\Delta\chi^2$ profiles are shown in Figure~\ref{fig:11} (Appendix C). The analysis demonstrates that most parameters remain within the $1\sigma$--$3\sigma$ range for small variations of order $\pm(3\text{--}5)\%$, indicating local stability of the fit. However, parameters such as $\beta_{\text{NS}}$ and $v_{\text{NS}}$ exhibit a comparatively steeper $\chi^2$ dependence, suggesting that the model is more sensitive to variations along these directions. Overall, the fit is well-behaved around the minimum and does not exhibit any pathological fine-tuning.

\subsection{Neutrinoless Double Beta Decay}

\begin{figure}[htbp]
\centering
\includegraphics[width=0.95\linewidth]{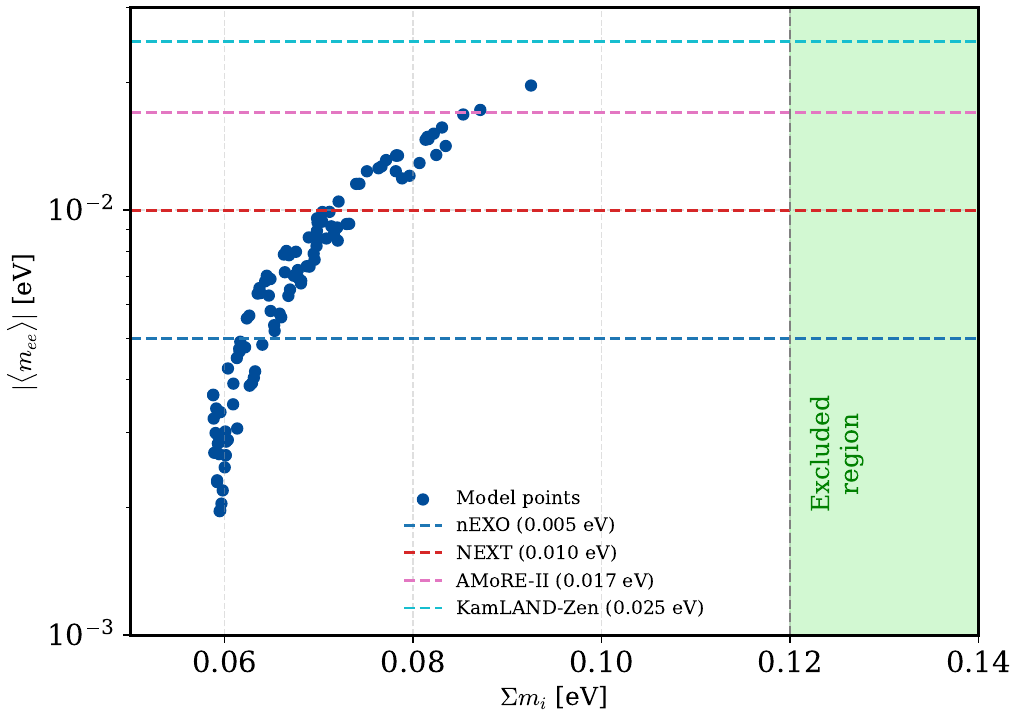}
\caption{Total neutrino mass $\sum m_\nu$ vs effective Majorana mass $|m_{ee}|$. }
\label{fig:mee_vs_sum}
\end{figure}

\begin{figure*}[t]
    \centering
    \begin{minipage}{0.49\linewidth}
        \centering
        \includegraphics[width=\linewidth]{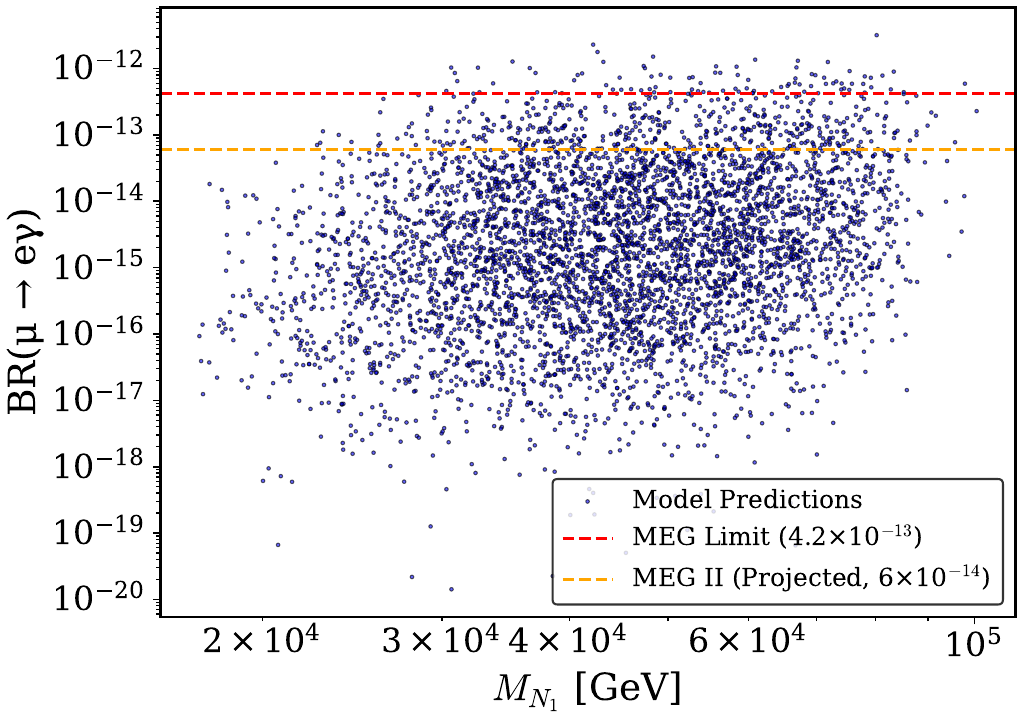}
    \end{minipage}
    \hfill
    \begin{minipage}{0.49\linewidth}
        \centering
        \includegraphics[width=\linewidth]{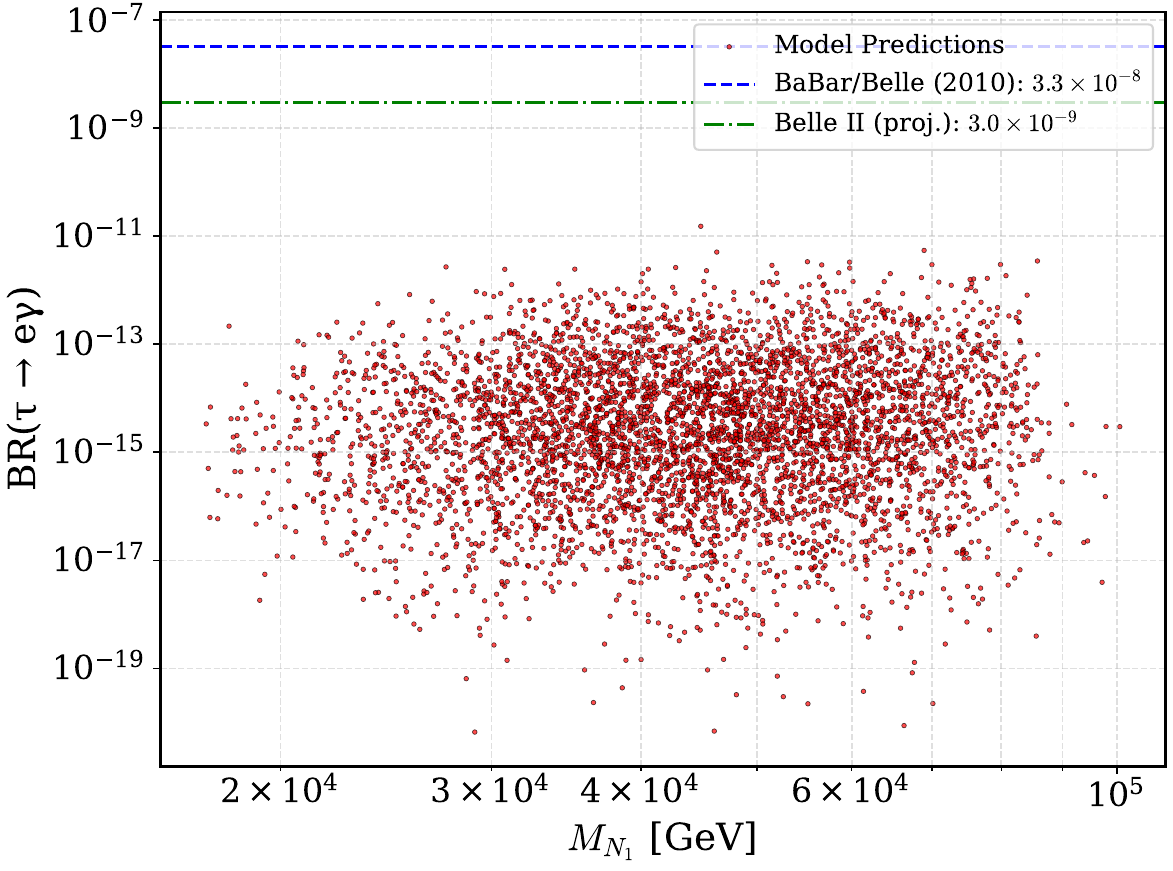}
    \end{minipage}
    
    \vspace{0.5cm}
    \begin{minipage}{0.49\linewidth}
        \centering
        \includegraphics[width=\linewidth]{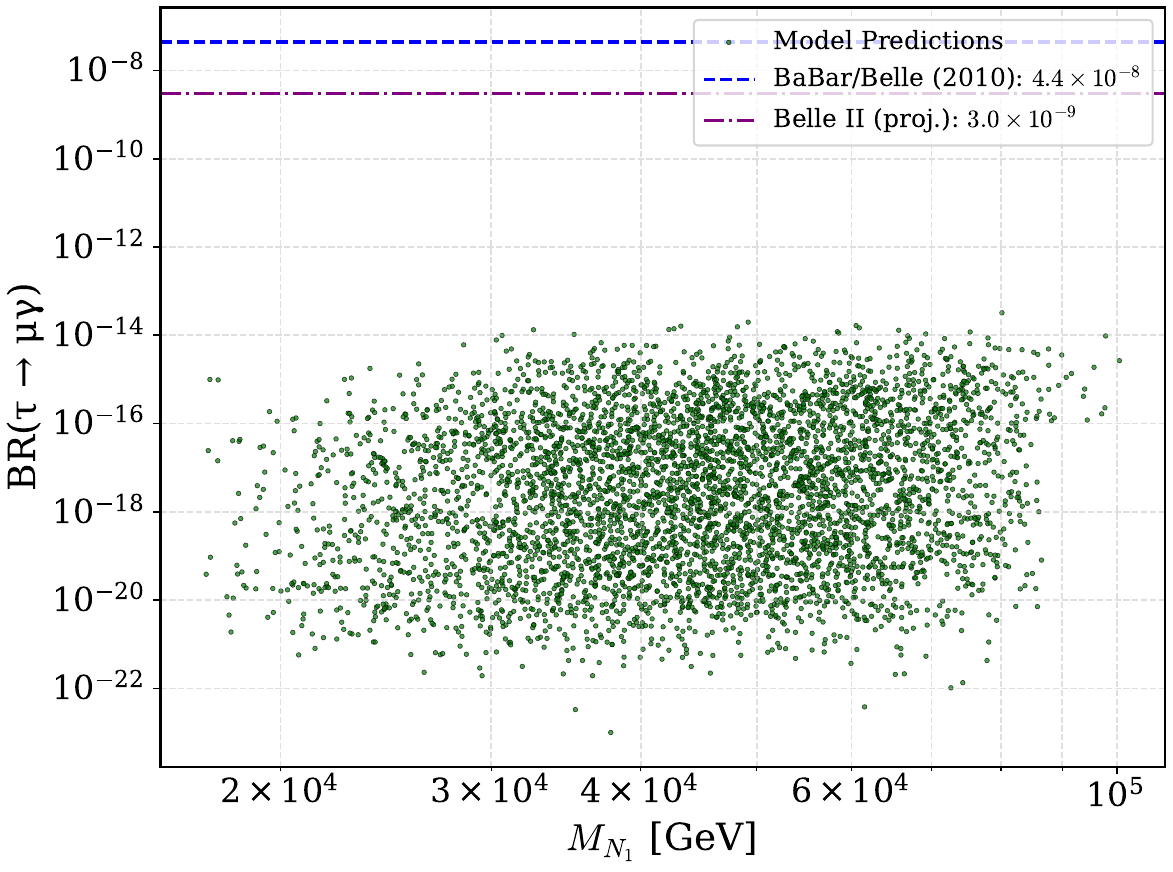}
    \end{minipage}
    
    \caption{$M_{N_1}$ dependence of $\mu\to e\gamma$, $\tau\to\mu\gamma$ and $\tau\to e\gamma$ branching ratios in the $S_4$ modular ISS(2,3) model. Dashed lines indicate current experimental bounds.}
    \label{fig:BR}
\end{figure*}

The neutrinoless double beta decay ($0\nu\beta\beta$) represents a crucial probe of lepton number violation and the Majorana nature of neutrinos. This hypothetical nuclear process would manifest as:

\begin{equation}
(Z,\, A) \rightarrow (Z + 2,\, A) + 2e^-
\label{eq:0nubb_process}
\end{equation}

The observation of this decay would establish neutrinos as Majorana particles while violating lepton number conservation by $\Delta L = 2$ units. The decay rate is governed by the effective Majorana mass:

\begin{equation}
|m_{ee}| = \left| m_1 c_{12}^2 c_{13}^2 + m_2 s_{12}^2 c_{13}^2 e^{i\alpha_{21}} + m_3 s_{13}^2 e^{i(\alpha_{31}-2\delta_{CP})} \right|
\label{eq:effective_mass}
\end{equation}
where $c_{ij} \equiv \cos\theta_{ij}$, $s_{ij} \equiv \sin\theta_{ij}$ and $\alpha_{21}$, $\alpha_{31}$, $\delta_{CP}$ are the CP-violating phases.

The model prediction for the effective Majorana mass $|m_{ee}|$ associated with neutrinoless double beta decay is displayed against $\sum m_\nu$ in Figure~\ref{fig:mee_vs_sum}.  According to the model,
\[
|m_{ee}| \in [0.002,\ 0.02]~\text{eV},
\]

which overlaps with the present experimental sensitivity and also under the projected limits of future experiments like nEXO\cite{nEXO:2021ujk}, NEXT\cite{NEXT:2023daz},  LEGEND~\cite{LEGEND2021},  KamLAND-Zen~\cite{KamLANDZen2016} and AMoRE-II\cite{Kim:2024ioo}. This makes $|m_{ee}|$ an important testable prediction of the model.

\subsection{Lepton Flavor Violation}

LFV refers to processes in which one type of charged lepton, such as a muon or tau, transforms into another without the emission of a corresponding neutrino to conserve flavor. In the minimal SM with massless neutrinos, such processes are strictly forbidden. Once neutrino masses and mixings are introduced, LFV becomes possible, but the predicted rates are so heavily suppressed (\(\text{BR}(\mu \to e\gamma) \sim 10^{-54}\)) that they are unobservable with current or foreseeable experimental precision. Consequently, any observation of LFV decays would constitute a clear and direct signal of physics beyond the SM. Among the most studied LFV processes are muon decays (e.g., $\mu \to e\gamma$) and tau decays (e.g., $\tau \to \mu\gamma$), because muons can be produced and controlled in large numbers in precision experiments, and taus being heavier are more sensitive to possible high-scale effects.

The effective mass matrix of neutrino is diagonalized via the unitary transformation \( U^\dagger \mathcal{M}_\nu U = \text{diag}(m_1, m_2,\\ m_3,      M_{N_1}, \ldots, M_{N_5}) \), where the mixing between light and heavy states gives rise to lepton flavor-violating effects. The most phenomenologically significant signatures arise in radiative decays \(\ell_i \to \ell_j \gamma\), whose branching ratios can be expressed as~\cite{Ilakovac:1995,Deppisch:2005,Forero:2011,Chekkal:2017,Chakraborty:2021azg}:
\begin{equation}
\text{BR}(\ell_i \to \ell_j \gamma) = 
\frac{\alpha_W^3 s_W^2}{256 \pi^2}\,
\frac{m_{\ell_i}^5}{M_W^4 \Gamma_{\ell_i}}
\left| \sum_{\alpha} U_{i\alpha} U_{j\alpha}^* \,
F\!\left(\frac{M_{N_\alpha}^2}{M_W^2}\right) \right|^2 ,
\label{eq:br_rad}
\end{equation}

where \(\alpha_W = g^2 / 4\pi\), \(s_W \equiv \sin \theta_W\), \(m_{\ell_i}\) and \(\Gamma_{\ell_i}\) are the charged lepton mass and total width, and \(m_{N_\alpha}\) are the heavy neutrino masses. The loop function contains the characteristic mass dependence,
\begin{equation}
F(x) = \frac{-2x^3 + 5x^2 - x}{4(1-x)^2} - \frac{3x^3}{2(1-x)^4}\,\ln x.
\label{eq:loop}
\end{equation}
Numerical inputs are taken from the Particle Data Group\cite{{PDG:2020}}, including \(\Gamma_\mu = 2.996 \times 10^{-19}\,\text{GeV}\) and \(\Gamma_\tau = 2.267 \times 10^{-12}\,\text{GeV}\).

Figure~\ref{fig:BR} displays the predicted branching ratios for the LFV channels $\mu\to e\gamma$, $\tau\to\mu\gamma$ and $\tau\to e\gamma$. We find that the branching ratios for $\tau \to e \gamma$ and $\tau \to \mu \gamma$ are approximately equal. This is not a generic prediction of the model but arises due to the chosen parameter values, which yield similar off-diagonal entries in $M_D^\dagger M_D$. The dashed horizontal line in the upper-left panel marks the MEG upper limit of $3.1\times10^{-13}$~\cite{MEG2016} for the muon decay, while the lines in the upper-right and lower panels show the current experimental limits from the BaBar collaboration, $\text{BR}(\tau \to \ell\gamma) < 3.3 \times 10^{-8}$~\cite{BaBar:2010}. Across the scanned parameter space, the model predictions lie comfortably below these limits, demonstrating consistency with current data. This compatibility supports the viability of the model and motivates further exploration in regions that may be probed by future experiments such as MEG~II~\cite{{Venturini:2024keu}} and Belle~II~\cite{BelleLFV2021}.

\section{Resonant Leptogenesis}
\label{sec:RL}
\subsection{Overview of the Mechanism}
To evaluate the model's consistency with cosmological observations, we examine whether it can account for the matter-antimatter imbalance observed in the universe via a leptogenesis mechanism.
 In the framework of the ISS(2,3) model, there are five heavy neutrino states: two of them are right-handed neutrinos ($N_1$, $N_2$), while the remaining three are sterile fermions ($S_i$).

These five states form one decoupled state and two quasi-Dirac pairings. Interestingly, a crucial aspect of resonant leptogenesis is that the mass splittings within these pairs are ordered by their decay widths. The decay of the lighter quasi-Dirac neutrino pair, occurring outside of equilibrium, serves as the source of lepton asymmetry in this framework. Through electroweak sphaleron transitions, this asymmetry is subsequently partially transformed into baryon asymmetry.

Although the heavier quasi-degenerate pair $(N_3, N_4)$ can in principle generate a lepton asymmetry, its contribution is strongly suppressed by both thermal and dynamical effects. At the relevant temperatures $T \sim 10^{4\text{--}5}\,\mathrm{GeV}$, these states are thermally inaccessible due to an exponential Boltzmann suppression $e^{-M_{3,4}/T}$, while any residual asymmetry is efficiently erased by lepton-number-violating washout from the lighter pair. Quantitatively, the corresponding washout parameters $K_{1,2} = \Gamma_{1,2}/H(T=M_{1,2}) \gg 1$ ensure that only the asymmetry from $(N_1, N_2)$ survives until sphaleron freeze-out. This feature aligns with the standard expectation that the lightest (or resonant) right-handed neutrino pair dominantly sources the baryon asymmetry in low-scale resonant leptogenesis~\cite{Blanchet:2006be,Blanchet:2012bk,Marciano:2024JHEP}.

\subsection{Generation of CP Asymmetry via Quasi-Degenerate Neutrino Decays}
\begin{figure*}[t]
\centering
\begin{minipage}{0.32\linewidth}
\centering
\includegraphics[width=0.9\linewidth]{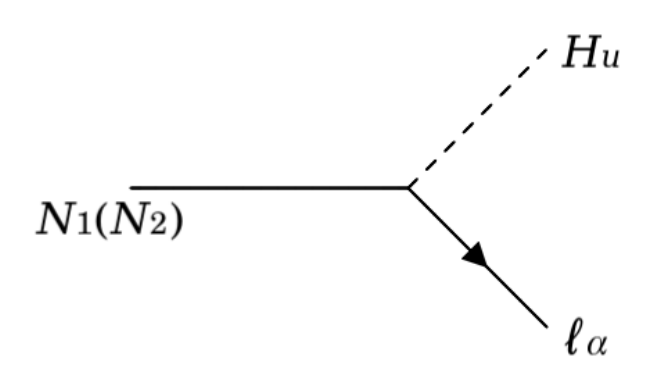}
\\(a)
\end{minipage}
\hfill
\begin{minipage}{0.32\linewidth}
\centering
\includegraphics[width=0.9\linewidth]{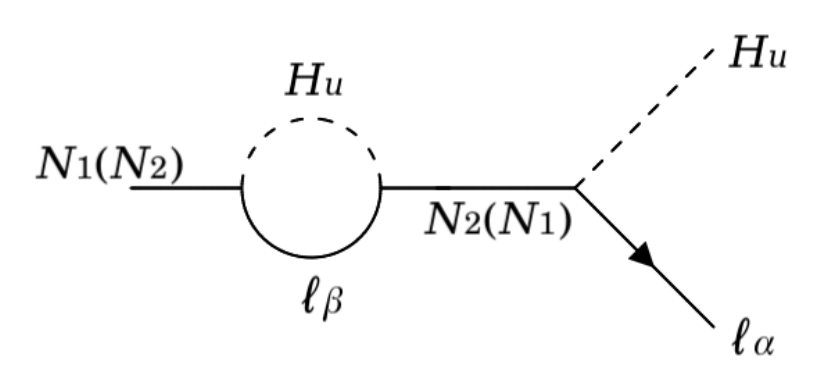}
\\(b)
\end{minipage}
\hfill
\begin{minipage}{0.32\linewidth}
\centering
\includegraphics[width=0.9\linewidth]{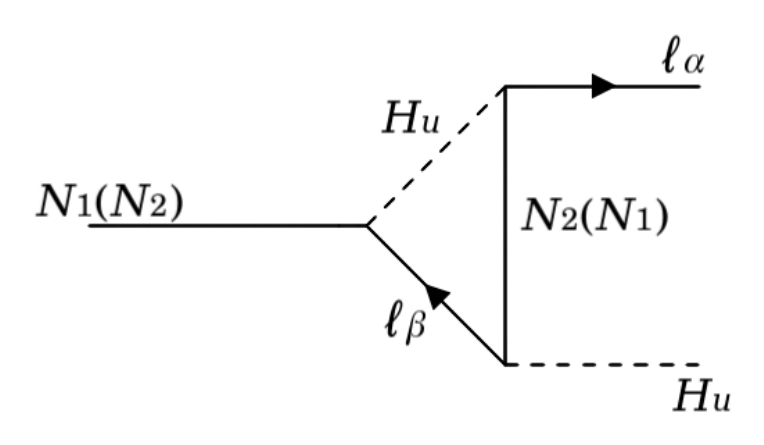}
\\(c)
\end{minipage}
\caption{Feynman diagrams contributing to the CP asymmetry through $N_1(N_2)$ decay: (a) tree-level, (b) vertex correction, and (c) self-energy correction.}
\label{fig:cp-decay}
\end{figure*}

Here, we analyze the CP asymmetry arising from the decays of two nearly degenerate right-handed neutrino states, labeled \( N_1 \) and \( N_2 \). These states emerge through the diagonalization of the lower \( 5 \times 5 \) submatrix of matrix $\mathcal{M}_\nu$ given in equation (\ref{eq:full_matrix}), which is expressed as:

\begin{equation}
M_H = 
\begin{pmatrix}
0 & M_{NS} \\
M_{NS}^T & M_S
\end{pmatrix}
\label{eq:MH}
\end{equation}

Diagonalization of $M_H$ via a unitary matrix $V$ yields four heavy Majorana states:

\begin{equation}
M_H^{\text{diag}} = V^T M_H V = \text{diag}(M_{N_1}, M_{N_2}, M_{N_3}, M_{N_4},  M_{N_5})
\end{equation}
These form two quasi-Dirac pairs: $(N_1, N_2)$ and $(N_3, N_4)$. In our setup, only the lighter pair $(N_1, N_2)$ contributes significantly to produce the lepton asymmetry due to their favorable mass splitting and suppressed washout. The complete $5 \times 5$ mass matrix is not easily tractable analytically, and we work with its numerical diagonalization for convenience. For the computation of the CP asymmetry, it is useful to choose a basis where the heavy neutrino mass matrix is diagonal, in which case the relevant part of the superpotential in equation(\ref{eq:2.5}) takes a simplified form:
 \begin{equation}
W_h = h_{i\alpha}N_i^c L_\alpha H_u  
\end{equation}
 The couplings \( h_{i\alpha} \) in the mass eigenstate basis are obtained from the flavor basis Yukawa couplings \( y_{i\alpha} \) (Yuk-awa coupling between $L_\alpha$ and $N_i$) through a unitary rotation defined by the matrix \( V \), as expressed below~\cite{Chakraborty:2021azg}:
\begin{align}
h_{1\alpha} &= V_{11}^* y_{1\alpha} + V_{12}^* y_{2\alpha} \\
h_{2\alpha} &= V_{21}^* y_{1\alpha} + V_{22}^* y_{2\alpha} \\
h_{3\alpha} &= V_{13}^* y_{1\alpha} + V_{23}^* y_{2\alpha} \\
h_{4\alpha} &= V_{14}^* y_{1\alpha} + V_{24}^* y_{2\alpha}
\end{align}

The CP asymmetry $\varepsilon_i$ is defined as~\cite{Chakraborty:2021azg,Garny:2013,Iso:2014}:

\begin{equation}
\varepsilon_i = \sum_{\alpha =e, \mu, \tau}  \frac{\left[{\Gamma(N_i \rightarrow \ell_\alpha H_u) - \Gamma(N_i \rightarrow \bar{\ell}_\alpha H_u^\dagger)}\right]}{\left[ \Gamma(N_i \rightarrow \ell_\alpha H_u) + \Gamma(N_i \rightarrow \bar{\ell}_\alpha H_u^\dagger) \right]}
\end{equation}

In the resonant leptogenesis regime, where $|M_{N_i} - M_{N_j}| \sim \Gamma_i$, the dominant contribution arises from self-energy effects and the asymmetry becomes~\cite{Chakraborty:2021azg}:

\begin{equation}
\varepsilon_i = \frac{1}{8\pi} \sum_{j \neq i} \frac{\text{Im} \left[ (h h^\dagger)^2_{ij} \right] f_{ij}}{(h h^\dagger)_{ii}}
\label{eq:cpasym}
\end{equation}

The self-energy loop function $f_{ij}$ is given by~\cite{Chakraborty:2021azg}:
\begin{equation}
f_{ij}^{\text{self}} = \frac{(M_{N_i}^2 - M_{N_j}^2) M_{N_i} M_{N_j}}{(M_{N_i}^2 - M_{N_j}^2)^2 + R_{ij}}
\label{eq:fij}
\end{equation}

with the regulator term:

\begin{equation}
R_{ij} = \left( M_{N_i} \Gamma_i + M_{N_j} \Gamma_j \right)^2
\label{eq:regulator}
\end{equation}

$N_i$'s overall decay width is~\cite{Chakraborty:2021azg}:

\begin{equation}
\Gamma_i = \frac{(h h^\dagger)_{ii} M_{N_i}}{8\pi}
\label{eq:decaywidth}
\end{equation}

As shown in~\cite{Chakraborty:2021azg}, the CP asymmetry for the lightest pair \( (N_1, N_2) \) is given by:

\begin{align}
\varepsilon_1 &= \frac{1}{8\pi (h h^\dagger)_{11}} \text{Im} \left[ (h h^\dagger)_{12}^2 f_{12} + (h h^\dagger)_{13}^2 f_{13} + (h h^\dagger)_{14}^2 f_{14} \right] \\
\varepsilon_2 &= \frac{1}{8\pi (h h^\dagger)_{22}} \text{Im} \left[ (h h^\dagger)_{21}^2 f_{21} + (h h^\dagger)_{23}^2 f_{23} + (h h^\dagger)_{24}^2 f_{24} \right]
\end{align}

The decay processes of these states lead to the generation of a net lepton asymmetry. However, this asymmetry can be partially washed out by lepton number violating scatterings and inverse decays, particularly if these processes remain efficient due to the presence of lighter sterile states or closely spaced mass eigenstates. The survival of the asymmetry depends on the interplay between the decay rate, mass splitting, and the strength of these washout processes. For instance, if the right-handed neutrinos \( N_{1} \) and \( N_{2} \) are lighter than \( N_{3} \) and \( N_{4} \), and have nearly degenerate masses (\( M_{N_1} \approx M_{N_2} \)), their out-of-equilibrium decays into SM leptons and Higgs doublets can produce significant CP asymmetries due to resonant enhancement. The associated CP asymmetry parameters \( \varepsilon_1 \) and \( \varepsilon_2 \), enter the Boltzmann equations that explains the generation and evolution of the lepton number asymmetry in the primordial universe.
\begin{figure*}[htbpt]
    \centering
    \begin{minipage}{0.48\linewidth}
        \centering
        \includegraphics[width=\linewidth]{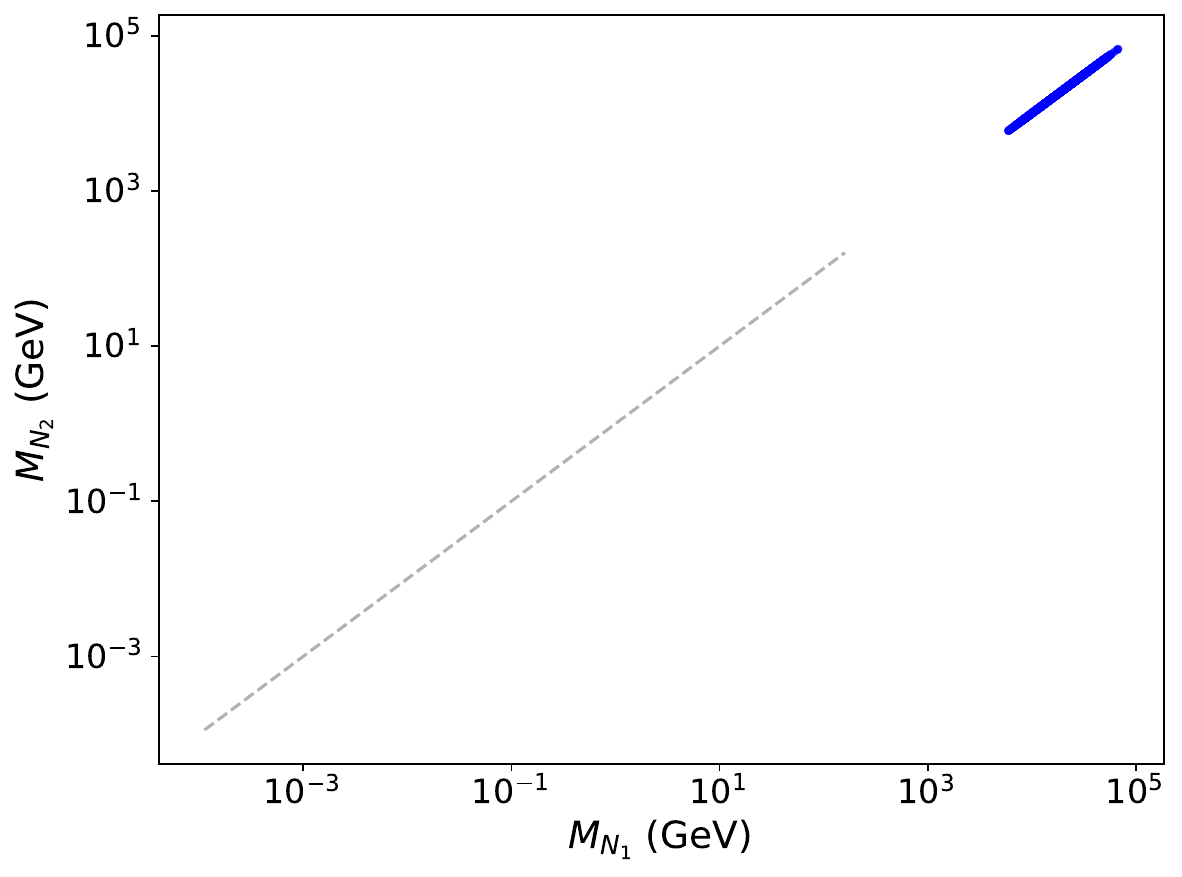}
    \end{minipage}
    \hfill
    \begin{minipage}{0.48\linewidth}
        \centering
        \includegraphics[width=\linewidth]{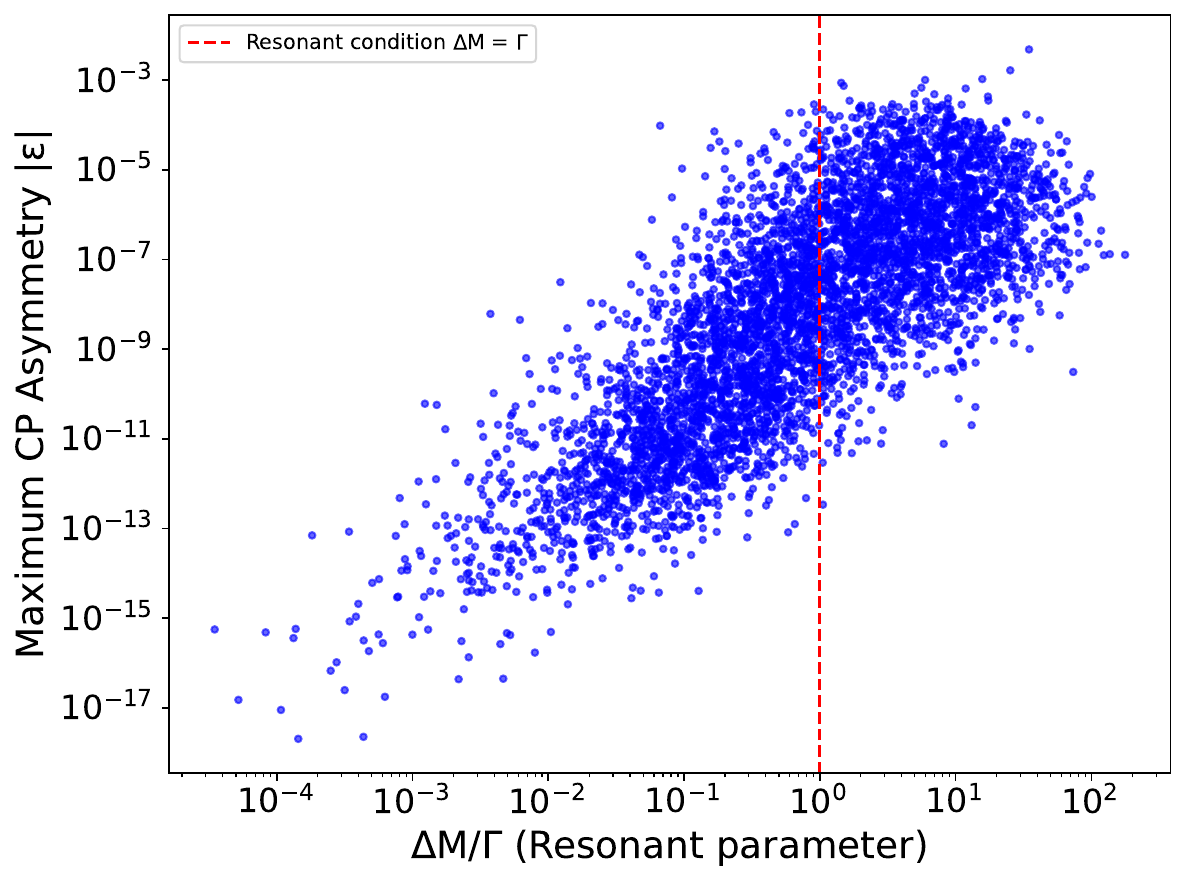}
    \end{minipage}
    \caption{
    Resonant leptogenesis analysis: 
    (left) $M_{N_1}$-$M_{N_2}$ mass correlation showing required near-degeneracy;
    (right) Resonance peak in $\varepsilon$ at $\Delta M/\Gamma \sim 1$.
    }
    \label{fig:leptogenesis_results}
\end{figure*}

\subsection{Boltzmann Equations and Evolution of Lepton Asymmetry}
We analyze how the number densities of \( N_1 \) and \( N_2 \) change over time to determine the resulting matter – antimatter asymmetry. This analysis is performed within a kinetic framework governed by differential equations that capture particle production, decay and asymmetry generation. For computational tractability, we maintain the approximation that SM degrees of freedom remain thermally coupled to the plasma throughout the studied cosmological period.

Let \( Y_{N_i} \) represent the comoving abundance of the heavy neutrino \( N_i \),  i.e. \( Y_{N_i} \equiv n_{N_i} / s \).
The evolution of $Y_{N_1}$, $Y_{N_2}$, and $Y_{B-L}$ is described by~\cite{Iso:2011}:

\begin{equation}
\frac{dY_{N_1}}{dz} = -\frac{z}{s H(M_{N_1})} 
\left( \frac{Y_{N_1}}{Y_{N_1}^{\text{eq}}} - 1 \right)\gamma_D^{(1)},
\end{equation}

\begin{equation}
\frac{dY_{N_2}}{dz}  = -\frac{z}{s H(M_{N_1})} 
\left( \frac{Y_{N_2}}{Y_{N_2}^{\text{eq}}} - 1 \right)\gamma_D^{(2)}, \end{equation}

\begin{equation}
\begin{split}
\frac{dY_{B-L}}{dz} = -\frac{z}{s H(M_{N_1})} 
\sum_{i=1}^2 \Bigg[ 
\varepsilon_i \left( \frac{Y_{N_i}}{Y_{N_i}^{\text{eq}}} - 1 \right)\gamma_D^{(i)} \\
+ \frac{Y_{B-L}}{2Y_\ell^{\text{eq}}}\,\gamma_D^{(i)}
+ \frac{Y_{B-L}}{Y_\ell^{\text{eq}}}\,\big(2\gamma_t + 2\gamma_s\big) 
\Bigg].
\end{split}
\end{equation}

Here, $z = M_{N_1}/T$ is a dimensionless time parameter, $H(M_{N_1})$ is the Hubble expansion rate at $T = M_{N_1}$ and $\gamma_D^{(i)}$ is the thermally averaged decay rate of $N_i$. The contributions to the total scattering rate are decomposed into their respective channels, denoted as $\gamma_s$ for the $s$ channel and $\gamma_t$ for the $t$ channel.

The Hubble expansion rate at \( T = M_{N_1} \) can be written as:

\begin{equation}
H(M_{N_1}) = \sqrt{\frac{4\pi^3 g_*}{45}} \frac{M_{N_1}^2}{M_{\text{Pl}}}
\end{equation}

With \( g_* \approx 110 \) and \( M_{\text{Pl}} = 1.22 \times 10^{19}~\text{GeV} \). 

Assuming equilibrium conditions, the number densities for the lepton and heavy neutrinos are~\cite{Chakraborty:2021azg}:

\begin{align}
Y_{N_i}^{\text{eq}}(z) &= \frac{45}{4\pi^4} \frac{g_{N}}{g_*} z^2 K_2(z), \\
Y_\ell^{\text{eq}} &= \frac{45 \zeta(3)}{2\pi^4} \frac{g_\ell}{g_*}
\end{align}

where $K_2(z)$ is the Bessel function of the second kind, and the interinsic degrees of freedom of the heavy neutrinos and leptons are $g_{N} = 2$ and $g_\ell = 2$, respectively. The lepton asymmetry generated through this mechanism undergoes conversion into matter-antimatter asymmetry via electroweak sphaleron processes. The electroweak sphaleron processes freeze out around \( T_{\text{sph}} \sim 150~\text{GeV} \), setting the baryon asymmetry \( Y_B \) is written as~\cite{Chakraborty:2021azg}:

\begin{equation}
Y_B = \left( \frac{8N_f + 4N_H}{22N_f + 13N_H} \right) Y_{B-L}(z_{\text{sph}})
\label{eq:baryon_asymmetry}
\end{equation}

Here, $Y_{B-L}(z_{\text{sph}})$ represents the solution to the Boltzmann equations evaluated at $z = z_{\text{sph}} = M_{N_1}/T_{\text{sph}}$, where $M_{N_1}$ is the mass of the lightest heavy neutrino. For the SM case considered here, we have $N_f = 3$ fermion families and $N_{H_u} = 1$ Higgs doublet.
In our numerical analysis, we focus on the decay and scattering processes mediated by the heavy neutrinos (see Appendix C).
\paragraph{Flavour considerations:}
In the present work, we have employed the single-flavour Boltzmann framework for simplicity. 
Although at temperatures $T \sim 10^{4\text{--}5}~\mathrm{GeV}$ all charged-lepton Yukawa interactions are in equilibrium, placing the model in the fully three-flavoured regime
\cite{Davidson:2008bu,Blanchet:2012bk}, the total asymmetry is nevertheless dominated by the $\mu$-flavour. 
This follows from the flavour projectors derived from the Dirac mass matrix,
$(K_{1e}, K_{1\mu}, K_{1\tau}) \\ \simeq (0.23, 0.71, 0.04)$,
showing that the $\mu$ and $e$ flavours govern the decay and washout processes, while the $\tau$ contribution remains subdominant. 
As a result, the single-flavour Boltzmann equations used in this work provide an excellent approximation to the total $B\!-\!L$ asymmetry. 
A full three-flavoured treatment would mainly lead to small numerical corrections without changing the qualitative behaviour of the predicted baryon asymmetry.

\section{Numerical Results for Resonant Leptogenesis}
\label{sec:NAL}

This part of the study is devoted to a numerical exploration of resonant leptogenesis, focusing on the thermal evolution of key dynamical variables. For the numerical analysis, we adopt three representative benchmark points defined by the following essential inputs: 
(i) a common heavy neutrino mass $M_{N_1} = 10^{4-5}~\text{GeV}$, with the second state taken nearly degenerate and separated by a relative mass splitting of order $10^{-8}-10^{-7}$; 
(ii) the Dirac mass matrix combination $M_D^\dagger M_D$, determined from low-energy neutrino data and of order $10^{-5}~\text{GeV}^2$; 
and (iii) the CP asymmetry parameter $\varepsilon$, varied in the range $5 \times 10^{-15}$--$1 \times 10^{-3}$. 
These three inputs determine the decay widths and asymmetries entering the Boltzmann equations, and provide the basis for the results shown in the following Figures.

 The left plot of Figure~\ref{fig:leptogenesis_results} establishes the required mass degeneracy between $M_{N_1}$ and $M_{N_2}$, with viable parameter space concentrated along the diagonal where $|M_{N_1}-M_{N_2}| \ll M_{N_1}$. This near-degeneracy is essential for the resonant enhancement of CP violation. The right panel of Figure~\ref{fig:leptogenesis_results} depicts the maximum CP asymmetry $\varepsilon$ as a function of the dimensionless ratio $\Delta M / \Gamma$, where $\Delta M = |M_{N_1} - M_{N_2}|$ is the mass splitting and $\Gamma$ is the average decay width. A clear broad maximum when the mass difference becomes comparable to the total decay width. This behavior originates from the regulator term in the self-energy contribution [see Eq.~\ref{eq:fij}], where the denominator

\[
(M_{N_i}^2 - M_{N_j}^2)^2 + (M_{N_i} \Gamma_i + M_{N_j} \Gamma_j)^2
\]

becomes minimal. In this limit, the self-energy diagram induces a resonant enhancement of CP violation. For nearly degenerate masses ($M_{N_i} \approx M_{N_j} \equiv M$), this condition yields the well-known resonance criterion for leptogenesis:

\begin{equation}
|M_{N_1} - M_{N_2}| \simeq \frac{\Gamma_1 + \Gamma_2}{2}.
\end{equation}

When this condition is satisfied, the self-energy diagram contributes maximally, leading to a significant enhancement of CP violation through quantum interference between the two heavy neutrino states. Together, these panels provide comprehensive evidence that our model achieves the necessary conditions for successful resonant leptogenesis.

We have also examined whether the time-dependent oscillatory behavior of the CP asymmetry, discussed in \cite{DeSimone:2007gol}, could affect our results. In the present inverse-seesaw framework, the heavy-neutrino pairs have masses in the range
$M_N \sim 10^{4\text{--}5}~\mathrm{GeV}$, with relative mass splittings
$\Delta M/M \sim 10^{-8}\text{--}10^{-7}$.
The corresponding decay widths are typically $\Gamma_N \simeq 10^{-5}\text{--}10^{-3}~\mathrm{GeV}$,
yielding $\Delta M/\Gamma_N \gtrsim 10^{1\text{--}3}$.
The associated washout parameter, defined as $K = \Gamma_N/H(M_N)$,
lies in the range $K \sim 10^{3\text{--}7}$,
which corresponds to the strong-washout, overdamped regime. Under these conditions, coherent oscillations of the CP asymmetry are exponentially suppressed and effectively average out on timescales much shorter than the heavy-neutrino decay time. Therefore, the constant-$\epsilon$ approximation employed in our Boltzmann analysis accurately captures the dynamics of lepton-asymmetry generation within the parameter region explored.

\begin{figure*}[t]
    \centering
    \begin{minipage}{0.48\linewidth}
        \centering
        \includegraphics[width=\linewidth]{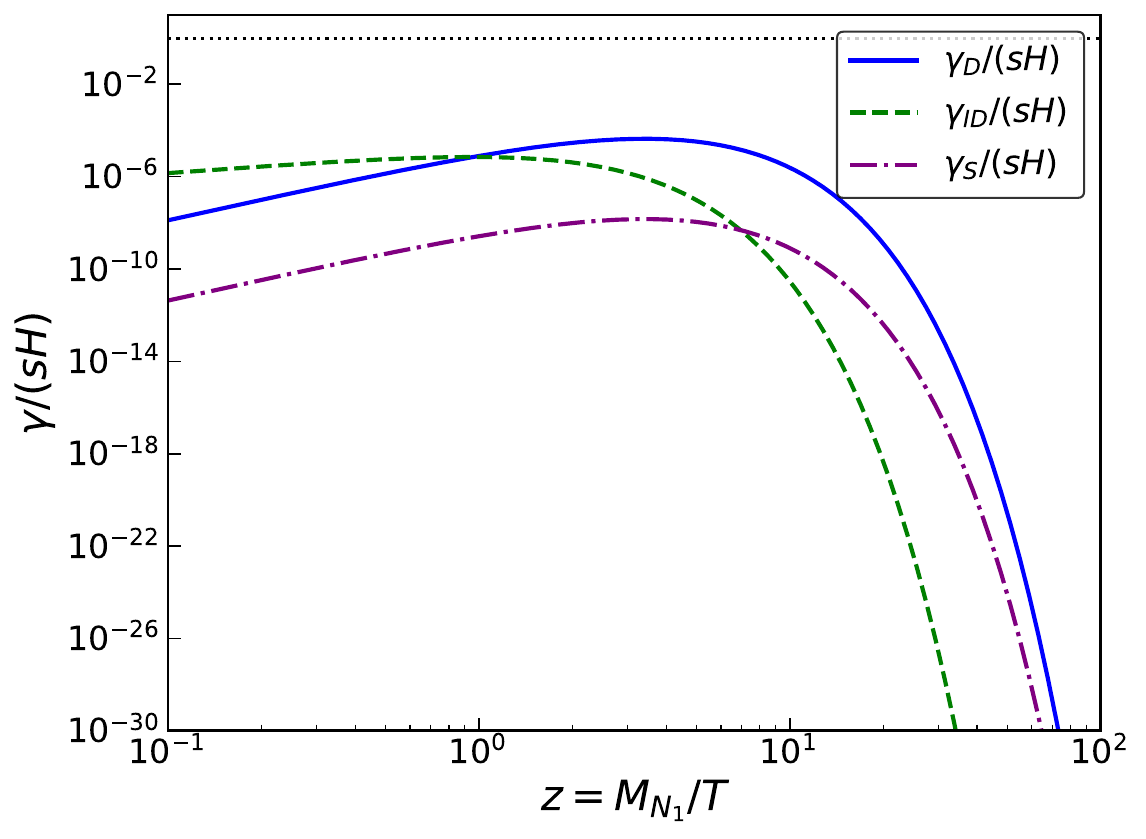}
    \end{minipage}
    \hfill
    \begin{minipage}{0.48\linewidth}
        \centering
        \includegraphics[width=\linewidth]{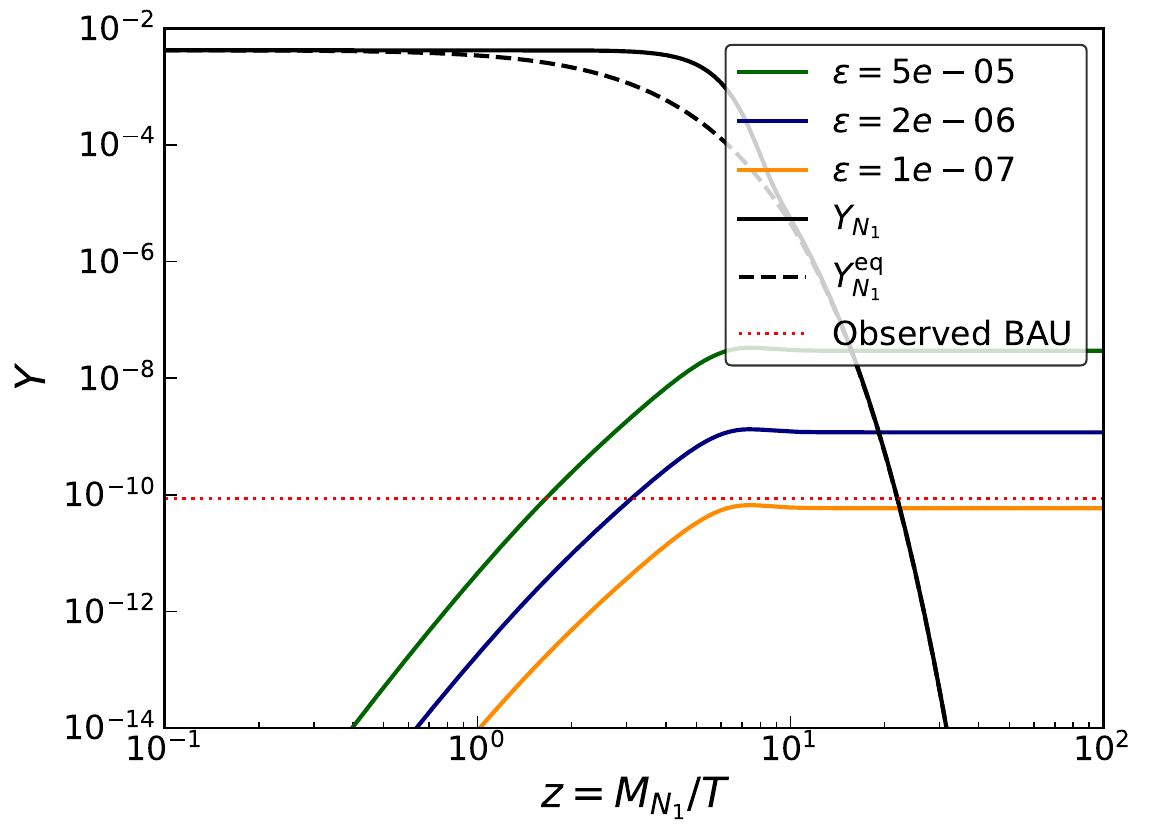}
    \end{minipage}
    \caption{Leptogenesis dynamics: Left panel shows thermal evolution of key interaction rates, and right panel displays $B\!-\!L$ asymmetry generation for different parameter sets.}

    \label{fig:YB}
\end{figure*}

The plot on the left of Figure \ref{fig:YB} presents the temperature evolution of different interaction rates $\gamma/(sH)$ relevant for leptogenesis, plotted against the parameter $z=M_{N_1}/T$, where $M_{N_1}$ is the mass of the heavy right-handed neutrino and $T$ is the temperature of the early universe. This variable serves as a time-like parameter: smaller values of $z$ correspond to earlier times (high temperatures), while larger $z$ values correspond to later times (cooler universe). This plot is crucial for visualizing whether and when the processes that generate lepton asymmetry occur out of thermal equilibrium, which is a core requirement for successful leptogenesis as per the Sakharov conditions.

\begin{figure}[t]
\centering
\begin{minipage}{0.48\linewidth}
\centering
\includegraphics[width=0.9\linewidth]{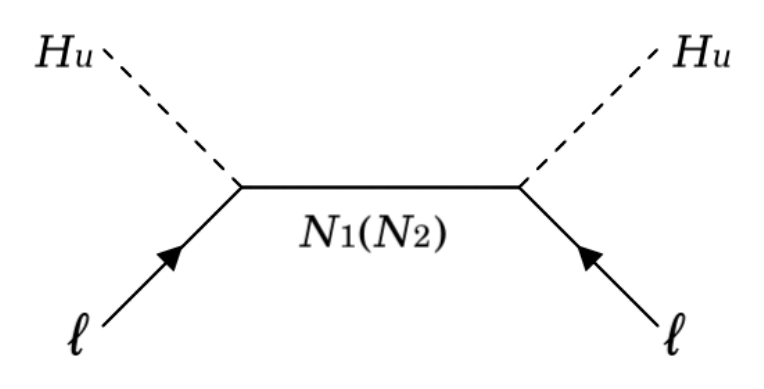}
\\ (a) $s$-channel
\end{minipage}
\hfill
\begin{minipage}{0.30\linewidth}
\centering
\includegraphics[width=0.9\linewidth]{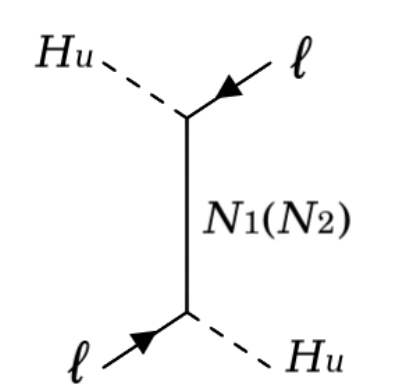}
\\ (b) $t$-channel
\end{minipage}
\caption{$\Delta L = 2$ scattering processes contributing to washout.}
\label{fig:scatter}
\end{figure}

The solid blue line denotes the decay contribution $\gamma_D/(sH)$, corresponding to the process $N_1 \to \ell H_u$. The green dashed curve represents the inverse decay $\gamma_{ID}/(sH)$, associated with $\ell H_u \to N_1$. The magenta dot–dashed curve shows the scattering contribution $\frac{\gamma_S}{(sH)}$. In the scattering sector as shown in Figure~\ref{fig:scatter} (\( \ell + H \leftrightarrow \ell + H,\ \) mediated by heavy neutrino \( N \) through \( s \)- and \( t \)-channels, both with \( \Delta L = 2 \)), the contribution of the $t$ channel remains present, and the contribution of the $s$ channel is strongly suppressed. The suppression originates from the fact that the $s$-channel diagram contains an intermediate heavy neutrino propagator, which can go on-shell. In this resonant configuration the contribution is already accounted for by the decay and inverse decay processes, and including it again would lead to double counting. Consequently, only the non-resonant remainder of the $s$-channel is kept, but this piece turns out to be negligible compared to the $t$-channel. Hence, the effective scattering rate shown in the plot is dominated by $t$-channel processes, with the $s$-channel effectively absent.

For reference, the horizontal dotted line at unity marks the equilibrium threshold $\gamma = sH$. In our case, we find that all interaction rates remain below the equilibrium line throughout the entire range of $z$. This ensures that the universe’s expansion dominates over the particle interaction rates, so that washout from inverse decays and scatterings is comparatively weak. As a result, once a lepton asymmetry is generated, it is not efficiently erased, fulfilling the out-of-equilibrium Sakharov condition.

The right panel of Figure~\ref{fig:YB} illustrates the evolution of the heavy neutrino abundance and the resulting lepton asymmetry as a function of $z = M_{N_1}/T$. The $y$-axis shows comoving number densities, i.e. abundances normalized to the entropy density, which is the standard way to describe particle populations in an expanding universe.

The black solid curve in Fig.~\ref{fig:YB} represents the comoving abundance of the lightest heavy neutrino $Y_{N_1}$, while the black dashed line denotes its equilibrium value $Y_{N_1}^{\mathrm{eq}}$.
At early times ($z \ll 1$), the heavy neutrinos are in thermal equilibrium with the plasma, as indicated by $Y_{N_1} \simeq Y_{N_1}^{\mathrm{eq}}$. As the temperature drops below $M_{N_1}$ ($z \gtrsim 1$), the equilibrium abundance becomes exponentially suppressed, and $Y_{N_1}$ departs from equilibrium, allowing the out-of-equilibrium decays of $N_1$ to generate a lepton asymmetry. For large washout parameters ($K \gg 1$), inverse decays remain efficient even after this initial departure, leading to a mild re-population of $N_1$ that causes $Y_{N_1}$ to temporarily approach $Y_{N_1}^{\mathrm{eq}}$ again before final freeze-out.

The coloured curves represent the baryon asymmetry $Y_B$ obtained for different benchmark values of the CP-asymmetry parameter $\varepsilon$. The horizontal red dotted line marks the observed baryon asymmetry of the Universe, $Y_B^{\rm obs} \simeq 8.6 \times 10^{-11}$. The generated asymmetry increases with $\varepsilon$; for the largest benchmark, the predicted $Y_B$ matches the observed value, while smaller $\varepsilon$ values yield proportionally lower asymmetries. 

The growth of $Y_B$ reflects the interplay between CP-violating decays, which source the asymmetry, and washout processes such as inverse decays and scatterings, which tend to erase it. In this setup, since all interaction rates remain below the equilibrium line in the left panel, the washout effects are relatively weak. As a result, the generated asymmetry is only mildly diluted and remains protected at late times. The differences between the benchmark curves correspond to the variation in $\varepsilon$: larger values produce a more rapid rise in $Y_B$, while smaller values lead to a slower accumulation of asymmetry.

\section{Conclusion}
\label{sec:C}
In summary, our modular \( S_4 \)-based inverse seesaw model with a (2,3) structure provides a predictive framework for neutrino masses and mixing. Through detailed numerical analysis, we demonstrate that the model successfully reproduces the observed neutrino oscillation parameters within the \( 3\sigma \) experimental ranges. The predicted values for mixing angles, CP-violating phases, and the absolute neutrino mass scale are tightly constrained. In particular, the Dirac CP phase is focused around  $\delta_{\rm CP} \approx \pm 90^\circ$ and the correlation between \( \delta_{\rm CP} \) and \( \sin^2\theta_{23} \) is consistent with NuFIT global data.

Beyond neutrino oscillation observables, the model offers testable predictions for low-energy phenomena. The effective Majorana mass \( |m_{ee}| \) falls below the current experimental bounds but lies within the sensitivity reach of upcoming \( 0\nu\beta\beta \) decay experiments, offering a promising avenue for future verification. Moreover, the model's projections for the rates of charged lepton flavor violating processes, such as \( \mu \rightarrow e\gamma \), lie well below current experimental bounds, thereby maintaining consistency with existing LFV constraints.

We also investigated the viability of resonant leptogenesis in this framework. The small split between the masses of the nearly degenerate heavy neutrinos fulfills the condition necessary for resonant enhancement of CP violation. Resonant enhancement of the CP asymmetry occurs as \( \Delta M \) approaches the scale of the decay width, with a peak observed around \( \Delta M/\Gamma \approx 1 \). The thermal evolution of the relevant interaction rates show that this framework successfully generates the observed matter-antimatter imbalance of the universe across a viable region of parameter space.

In summary, this modular \( S_4 \)-based ISS(2,3) framework offers a unified explanation for neutrino masses, CP violation, rare lepton decays, and the baryon asymmetry of the universe. Future experimental data from LFV searches and \( 0\nu\beta\beta \) decay experiments, as well as more precise neutrino oscillation measurements, will be critical in further testing and constraining this model.

\appendix
\section*{Appendix A: Modular Symmetry}
\label{app:A}
\addcontentsline{toc}{section}{Appendix A: Modular Symmetry}
\renewcommand{\theequation}{A.\arabic{equation}}
Recently, modular symmetry has emerged as a valuable technique for neutrino mass and mixing modeling.  By connecting Yukawa couplings to unique mathematical functions known as modular forms, it provides an alternative method for understanding flavor symmetries.  These functions depend on the modulus \( \tau \), a complex variable confined to the upper-half plane (\( \text{Im}(\tau) > 0 \)), which transforms non-trivially under modular symmetry according to:

\[
\tau \rightarrow \frac{a\tau + b}{c\tau + d},
\]
Here, \( a, b, c, d \) are integers constrained by the relation \( ad - bc = 1 \), which defines elements of the special linear group \( \mathrm{SL}(2, \mathbb{Z}) \), commonly referred to as the modular group. Subgroups of this group, known as modular groups of level \( N \) and denoted \( \Gamma(N) \), are formed by imposing congruence relations modulo \( N \).

There are two basic processes that create the modular group:
\begin{align}
S: &\quad \tau \mapsto -\frac{1}{\tau}, \\
T: &\quad \tau \mapsto \tau + 1.
\end{align}

At low modular levels \( N \), the finite modular groups \( \Gamma_N = \overline{\Gamma}(1)/\overline{\Gamma}(N) \) correspond to well-known discrete groups that frequently appear in flavor model building:
\[
\Gamma_2 \cong S_3,\quad \Gamma_3 \cong A_4,\quad \Gamma_4 \cong S_4,\quad \Gamma_5 \cong A_5.
\]

In this work, we focus on \( \Gamma_4 \simeq S_4 \), which serves as the underlying flavor symmetry. In this construction, the Yukawa interactions are encoded through modular forms, which carry non-trivial transformation properties under the action of the modular symmetry group. Depending on the theory's field content, these couplings are arranged into irreducible representations like singlets, doublets and triplets. The number of available modular forms is determined by both the modular weight and the level \( N \) of the group. The distribution of independent modular forms as a function of level and weight is summarized in Table~\ref{tab:modular_forms_simple}.

\begin{table}[htbp]
\centering
\caption{Number of modular forms of weight \( 2k \) for selected groups.}
\begin{tabular}{|c|c|c|}
\hline
Level \( N \) & Group & Number of forms (weight \( 2k \)) \\
\hline
2 & \( S_3 \) & \( k + 1 \) \\
3 & \( A_4 \) & \( 2k + 1 \) \\
4 & \( S_4 \) & \( 4k + 1 \) \\
5 & \( A_5 \) & \( 10k + 1 \) \\
\hline
\end{tabular}
\label{tab:modular_forms_simple}
\end{table}

By choosing a suitable value for \( \tau \), we can determine the structure of Yukawa couplings and, as a result, predict neutrino masses and mixing angles. This makes modular symmetry a powerful and elegant way to build flavor models in particle physics.

\section*{Appendix B: Modular Yukawa Structures under \( S_4 \)}
\label{app:B}
\addcontentsline{toc}{section}{Appendix B: Modular Yukawa Structures under \( S_4 \)}

This $S_4$ group is composed of four objects permuted.  $1$, $1'$, $2$, $3$, and $3'$ are its five irreducible representations, and it has two generators that meet the following criteria:
\begin{equation}
S^2 = T^4 = (ST)^3 = I. 
\end{equation}

For convenience, we present a collection of essential ingredients used in model construction, including Clebsch–Gordan decompositions, \( q \)-expansions of modular forms at low weights and the corresponding multiplet structures, all formulated within the symmetric basis. This collection is intended to support the reader in model-building applications.

\subsection*{1. Products involving \(1'\)}
\begin{align}
1' \otimes 1' &= 1, \quad 1 \sim \alpha_1 \beta_1, \\
1' \otimes 2 &= 2, \quad 2 \sim \begin{pmatrix} -\alpha_1 \beta_2 \\ \alpha_1 \beta_1 \end{pmatrix}, \\
1' \otimes 3 &= 3', \quad 3' \sim \begin{pmatrix} \alpha_1 \beta_1 \\ \alpha_1 \beta_2 \\ \alpha_1 \beta_3 \end{pmatrix}, \\
1' \otimes 3' &= 3, \quad 3 \sim \begin{pmatrix} \alpha_1 \beta_1 \\ \alpha_1 \beta_2 \\ \alpha_1 \beta_3 \end{pmatrix}.
\end{align}

\subsection*{2. Product \(2 \otimes 2\)}
\begin{equation}
2 \otimes 2 = 1 \oplus 1' \oplus 2,
\end{equation}
\small{
\begin{align*}
1 \sim \alpha_1 \beta_1 + \alpha_2 \beta_2, \,\,\
1' \sim \alpha_1 \beta_2 - \alpha_2 \beta_1, \,\,\
2 \sim \begin{pmatrix} \alpha_2 \beta_2 - \alpha_1 \beta_1 \\ \alpha_1 \beta_2 + \alpha_2 \beta_1 \end{pmatrix}.
\end{align*}}

\subsection*{3. Products \(2 \otimes 3\) }
\begin{equation}
2 \otimes 3 = 3 \oplus 3',
\end{equation}
\small{
\begin{align*}
3 \sim \begin{pmatrix} \alpha_1 \beta_1 \\ \frac{1}{2}(\sqrt{3} \alpha_2 \beta_3 - \alpha_1 \beta_2) \\ \frac{1}{2}(\sqrt{3} \alpha_2 \beta_2 - \alpha_1 \beta_3) \end{pmatrix}, \quad
3' \sim \begin{pmatrix} -\alpha_2 \beta_1 \\ \frac{1}{2}(\sqrt{3} \alpha_1 \beta_3 + \alpha_2 \beta_2) \\ \frac{1}{2}(\sqrt{3} \alpha_1 \beta_2 + \alpha_2 \beta_3) \end{pmatrix}.
\end{align*}}

\subsection*{4. Products \(2 \otimes 3'\)}
\begin{equation}
2 \otimes 3' = 3 \oplus 3',
\end{equation}
\small{
\begin{align*}
3 \sim \begin{pmatrix} -\alpha_2 \beta_1 \\ \frac{1}{2}(\sqrt{3} \alpha_1 \beta_3 + \alpha_2 \beta_2) \\ \frac{1}{2}(\sqrt{3} \alpha_1 \beta_2 + \alpha_2 \beta_3) \end{pmatrix},
3' \sim \begin{pmatrix} \alpha_1 \beta_1 \\ \frac{1}{2}(\sqrt{3} \alpha_2 \beta_3 - \alpha_1 \beta_2) \\ \frac{1}{2}(\sqrt{3} \alpha_2 \beta_2 - \alpha_1 \beta_3) \end{pmatrix}.
\end{align*}}

\subsection*{5. Products \(3 \otimes 3\) and \(3' \otimes 3'\)}
\begin{equation}
3 \otimes 3 = 3' \otimes 3' = 1 \oplus 2 \oplus 3 \oplus 3',
\end{equation}
\begin{align*}
1 \sim \alpha_1 \beta_1 + \alpha_2 \beta_3 + \alpha_3 \beta_2, 
\end{align*}
\begin{align*}
2 \sim \begin{pmatrix} \alpha_1 \beta_1 - \frac{1}{2}(\alpha_2 \beta_3 + \alpha_3 \beta_2) \\ \frac{\sqrt{3}}{2}(\alpha_2 \beta_2 + \alpha_3 \beta_3) \end{pmatrix}, 
\end{align*}
\begin{align*}
3 \sim \begin{pmatrix} \alpha_3 \beta_3 - \alpha_2 \beta_2 \\ \alpha_1 \beta_3 + \alpha_3 \beta_1 \\ -\alpha_1 \beta_2 - \alpha_2 \beta_1 \end{pmatrix}, \quad
3' \sim \begin{pmatrix} \alpha_3 \beta_2 - \alpha_2 \beta_3 \\ \alpha_2 \beta_1 - \alpha_1 \beta_2 \\ \alpha_1 \beta_3 - \alpha_3 \beta_1 \end{pmatrix}.
\end{align*}

\subsection*{6. Product \(3 \otimes 3'\)}
\begin{equation}
3 \otimes 3' = 1' \oplus 2 \oplus 3 \oplus 3',
\end{equation}
\begin{align*}
1' \sim \alpha_1 \beta_1 + \alpha_2 \beta_3 + \alpha_3 \beta_2,
\end{align*} 
\begin{align*}
2 \sim \begin{pmatrix} \frac{\sqrt{3}}{2}(\alpha_2 \beta_2 + \alpha_3 \beta_3) \\ -\alpha_1 \beta_1 + \frac{1}{2}(\alpha_2 \beta_3 + \alpha_3 \beta_2) \end{pmatrix}, 
\end{align*}
\begin{align*}
3 \sim \begin{pmatrix} \alpha_3 \beta_2 - \alpha_2 \beta_3 \\ \alpha_2 \beta_1 - \alpha_1 \beta_2 \\ \alpha_1 \beta_3 - \alpha_3 \beta_1 \end{pmatrix}, \quad
3' \sim \begin{pmatrix} \alpha_3 \beta_3 - \alpha_2 \beta_2 \\ \alpha_1 \beta_3 + \alpha_3 \beta_1 \\ -\alpha_1 \beta_2 - \alpha_2 \beta_1 \end{pmatrix}.
\end{align*}

The modular forms of minimal weight can be represented through their \( q \)-expansion, with the corresponding basis functions taking the following explicit form:

{\small
\begin{equation}
\begin{split}
Y_1 = -3\pi \bigl( \tfrac{1}{8} + 3q + 3q^2 + 12q^3 + 3q^4 + 18q^5 + 12q^6 \\ + 24q^7 + 3q^8 + 39q^9 \bigr),
\end{split}
\end{equation}
\begin{equation}
\begin{split}
Y_2 = 3\sqrt{3}\pi q^{1/2} \bigl( 1 + 4q + 6q^2 + 8q^3 + 13q^4 + 12q^5 + 14q^6 \\+ 24q^7 + 18q^8 + 20q^9 \bigr),
\end{split}
\end{equation}
\begin{equation}
\begin{split}
Y_3 = \pi \bigl( \tfrac{1}{4} - 2q + 6q^2 - 8q^3 + 6q^4 - 12q^5 + 24q^6 - 16q^7 \\+ 6q^8 - 26q^9 + 38q^{10} \bigr),
\end{split}
\end{equation}
\begin{equation}
\begin{split}
Y_4 = -\sqrt{2}\pi q^{1/4} \bigl( 1 + 6q + 13q^2 + 14q^3 + 18q^4 + 32q^5 + 31q^6 \\+ 30q^7 + 48q^8 + 38q^9 \bigr),
\end{split}
\end{equation}
\begin{equation}
\begin{split}
Y_5 = -4\sqrt{2}\pi q^{3/4} \bigl( 1 + 2q + 3q^2 + 6q^3 + 5q^4 + 6q^5 + 10q^6 \\+ 8q^7 + 12q^8 + 14q^9 \bigr).
\end{split}
\end{equation}
}

where $q \equiv e^{i2\pi\tau}$. The multiplets in modular form with the lowest weight are:
\begin{equation}
Y_2 = \begin{pmatrix} Y_1 \\ Y_2 \end{pmatrix}, \quad Y_{3'} = \begin{pmatrix} Y_3 \\ Y_4 \\ Y_5 \end{pmatrix}. 
\end{equation}

The modular-form multiplets at weight four are:
\begin{align}
Y^{(4)}_1 &= Y^2_1 + Y^2_2, \quad Y^{(4)}_2 = \begin{pmatrix} Y^2_2 - Y^2_1 \\ 2Y_1 Y_2 \end{pmatrix},  \\
Y^{(4)}_3 &= \begin{pmatrix} -2Y_2 Y_3 \\ \sqrt{3}Y_1 Y_5 + Y_2 Y_4 \\ \sqrt{3}Y_1 Y_4 + Y_2 Y_5 \end{pmatrix},\quad Y^{(4)}_{3'} = \begin{pmatrix} 2Y_1 Y_3 \\ \sqrt{3}Y_2 Y_5 - Y_1 Y_4 \\ \sqrt{3}Y_2 Y_4 - Y_1 Y_5 \end{pmatrix}, 
\end{align}
where we clearly mention the modular weight using superscript ``$(4)$''. The group-theoretical structures, generator matrices, and modular form expansions presented here are adapted from Ref.~\cite{Novichkov:2019sqb,Zhang:2021}.

\section*{Appendix C: Decay and Scattering Rates}
\label{app:C}
\addcontentsline{toc}{section}{Appendix C: Decay and Scattering Rates}
\subsection*{Decay Rates}

The decay rate is given by~\cite{Plumacher:1996hd}:
\begin{equation}
\gamma_D \equiv \gamma^{\text{eq}}(N \to i + j + \cdots) = s Y_{N}^{\text{eq}} \Gamma_D \frac{K_1(z)}{K_2(z)}
\end{equation}
where:
\begin{itemize}
    \item $\Gamma_D = \Gamma_N \frac{K_1(z)}{K_2(z)}$ is the thermal decay width
    \item The tree-level decay width is:
    \begin{equation}
        \Gamma_{N_i} = \frac{\alpha}{\sin^2\theta_W} \frac{M_{N_i}(M_D^\dagger M_D)_{ii}}{4M_W^2}
    \end{equation}
In this expression, $\theta_W$ is the angle associated with electroweak mixing \\
and $M_W$ denotes the W boson mass.
\end{itemize}

\subsection*{Thermal averaged scattering cross sections:}
The reaction density for $N + a \leftrightarrow i + j + \cdots$ is~\cite{Plumacher:1996hd}:
\begin{equation}
\gamma^{\text{eq}} = \frac{T}{64\pi^4} \int_{s_{\text{min}}}^\infty \mathrm{d}s \hat{\sigma}(s) \sqrt{s} K_1\left(\frac{\sqrt{s}}{T}\right)
\end{equation}
where:
\begin{itemize}
    \item $s_{\text{min}} = \max[(M_N + M_a)^2, (M_i + M_j)^2]$
    \item The relationship between the total and reduced cross-sections is given by:
    \begin{equation}
        \hat{\sigma}(s) = \frac{8}{s} [(p_N \cdot p_a)^2 - M_N^2 M_a^2] \sigma(s)
    \end{equation}
\end{itemize}

\subsection*{Scattering Processes via Heavy Neutrino Exchange:}

For the $s$ and $t$ channel contributions mediated by the right-handed neutrinos $N_i$, the reduced cross sections are provided by~\cite{Chakraborty:2021azg}:

\paragraph{$s$-channel:}
{\footnotesize
\begin{multline}
\hat{\sigma}_{N,s}(s) =
\frac{\alpha^2}{\sin^4\theta_W}\,
\frac{2\pi}{M_W^4}\,
\frac{1}{x}
\Bigg\{
\sum_{j=1}^2 a_j (M_D^\dagger M_D)_{jj}^2
\Bigg[
\frac{x}{a_j} + \frac{2x}{D_j(x)} \\+ \frac{x^2}{2D_j(x)^2}
-\left(1+\frac{2x+a_j}{D_j(x)}\right)
\ln\!\biggl(\frac{x+a_j}{a_j}\biggr)
\Bigg]  
+\, 2\sqrt{a_1 a_2}\,
\Re\!\bigl[(M_D^\dagger M_D)_{12}^2\bigr]\\
\Bigg[
\frac{x}{D_1(x)} + \frac{x}{D_2(x)} 
+ \frac{x^2}{2D_1(x)D_2(x)} 
-\frac{(x+a_1)(x+a_1-2a_2)}{D_2(x)(a_1-a_2)}\\
\ln\!\biggl(\frac{x+a_1}{a_1}\biggr)
-\frac{(x+a_2)(x+a_2-2a_1)}{D_1(x)(a_2-a_1)}
\ln\!\biggl(\frac{x+a_2}{a_2}\biggr)
\Bigg]
\Bigg\}.
\end{multline}
}

with
\begin{align}
x = \frac{s}{M_{N_1}^2}, \quad 
a_i = \frac{M_{N_i}^2}{M_{N_1}^2}, \quad 
\end{align}
\begin{align}
\frac{1}{D_j(x)} = \frac{x-a_j}{(x-a_j)^2 + a_j c_j}, \quad
c_j = \left(\frac{\Gamma_{N_j}}{M_{N_1}}\right)^2.
\end{align}

\paragraph{$t$-channel:}
{\footnotesize
\begin{multline}
\hat{\sigma}_{N,t}(s) =
\frac{2\pi \alpha^2}{M_W^4 \sin^4\theta_W}
\Bigg\{
\sum_{j=1}^2 a_j (M_D^\dagger M_D)_{jj}^2
\Bigg[
\frac{1}{2a_j}\frac{x}{x+a_j}
\\+ \frac{1}{x+2a_j}\ln\!\biggl(\frac{x+a_j}{a_j}\biggr)
\Bigg] 
+\, \Re\!\bigl[(M_D^\dagger M_D)_{12}^2\bigr]\,
\frac{\sqrt{a_1 a_2}}{(a_1-a_2)(x+a_1+a_2)}\\
\Bigg[
(x+2a_1)\ln\!\biggl(\frac{x+a_2}{a_2}\biggr)
-(x+2a_2)\ln\!\biggl(\frac{x+a_1}{a_1}\biggr)
\Bigg]
\Bigg\}.
\end{multline}
}

\subsubsection*{$\chi^2$ Sensitivity Around the Best-Fit Point}

\begin{figure}[htbp]
    \centering
    \includegraphics[width=\columnwidth]{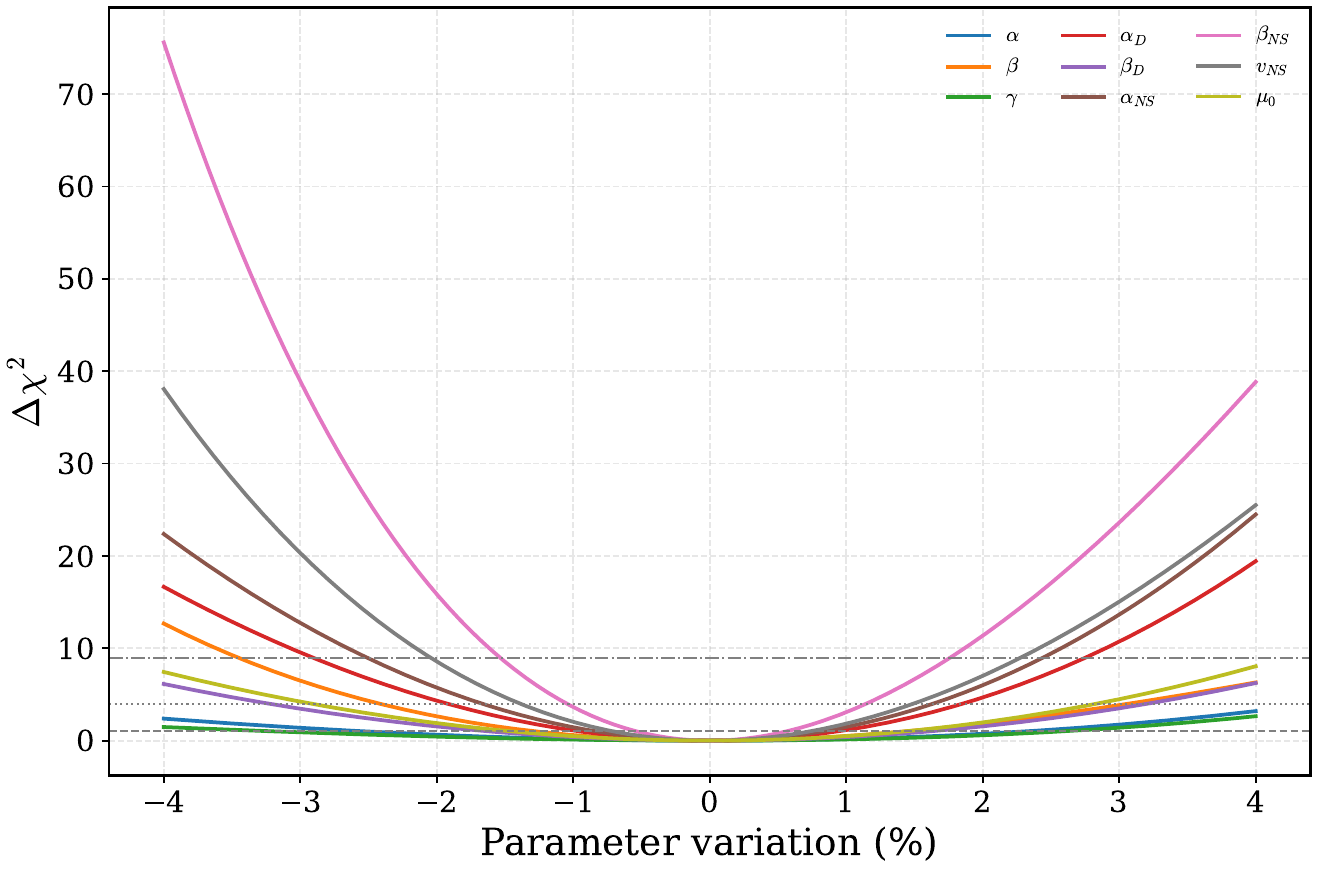}
    \caption{Variation of the total $\chi^2$ as a function of key model parameters around the best-fit point.}
    \label{fig:11}
\end{figure}

To test the local stability of the obtained solution, we varied each model parameter individually around the best-fit point while keeping the remaining parameters fixed. 
The resulting dependence of $\Delta\chi^2$ on the parameter variation is shown in Fig.~\ref{fig:11}. 
The dashed horizontal lines correspond to $\Delta\chi^2 = 1$, $4$, and $9$, which represent the $1\sigma$, $2\sigma$, and $3\sigma$ confidence levels for a single degree of freedom.

\bibliographystyle{unsrt}

\end{document}